# Supercritical fluid of quantum electrons in three-dimensional superconducting fullerides

Xinying Li[1,2,3], Yuki Matsuda[4], Chongli Yang[1], Huaxue Zhou[1], Takuma Ogasawara[1], Qin Wang[1], Liguo Zhang[1], Hidekazu Shimotani[4], Hailiang Xia[5], Khuong K. Huynh[6], Panagiotis Kotetes[1]*, Satoshi Heguri[7,8]*, Katsumi Tanigaki[1]*

[1]Beijing Academy of Quantum Information Sciences (BAQIS); Beijing, 100193, China
[2] Beijing National Laboratory for Condensed Matter Physics, Institute of Physics, Chinese Academy of Sciences; Beijing, 100190, China
[3]University of Chinese Academy of Sciences; Beijing, 100190, China
[4]Advanced Institute of Materials Research (AIMR), Tohoku University; Sendai, Japan
[5] Beijing Huihaoyu Intelligent Technology Co., Ltd.; Beijing, China.
[6]Department of Chemistry, Aarhus University; Aarhus, Denmark.
[7]Osaka Institute of Technology; Osaka, Japan
[8]Institute of Physics; Zagreb, Croatia.

## Abstract

The supercritical fluid (SCF) of quantum electrons at the Mott metal-insulator transition without symmetry breaking is one of the most elusive phenomena in strongly correlated electron physics. Prior studies of Cr-doped $V_2O_3$ and organic Mott systems reported discrepant critical exponents. A key limitation is that the scaling analysis relies on a single experimental observable, leaving the roles of phase coexistence, inhomogeneity, and percolation unaddressed. Here we report the first experimental identification of a thermodynamically equilibrated SCF phase and its associated Mott endpoint in the three-dimensional superconducting fullerides $Cs_xRb_{3-x}C_{60}$, using two independent probes of electrical conductivity and magnetic susceptibility, which reveal two distinct metal-insulator transition lines converging at a single Mott endpoint. A hypothesis-free two-particle analysis of magnetic susceptibilities yields a metal-insulator coexisting SCF by exhibiting the maximum two-phase mixing entropy, in agreement with a picture of a thermodynamically equilibrated Widom line. Simultaneously, conductivity scaling yields a critical exponent in the regime of quantum critical predictions. Our new dual-probe approach provides a unified microscopic picture of the Mott SCF with a characteristic length scale below current diffraction resolution, in addition to a new interpretation on the origin of superconducting $T_c$-dome.

## Main

In classical particles of atomic elements or molecules accommodated in a container, gas (G) and liquid (L) crossover without symmetry breaking above a finite critical temperature, as shown in Fig. 1(a). Such unique G-L fluids in a thermodynamically equilibrium state are called supercritical fluids (SCFs). The understanding of SCFs remains a long-standing and vital scientific topic to date[1-3]. Motivated by the SCFs of classical particles, an intriguing scientific question has been raised for the metal (M)-insulator (I) transition (MIT) of quantum particles of electrons in a crystal cell, as shown in Fig. 1(b). In solid-state physics, the first pivotal experiments were performed on a Mott material, Cr-doped $V_2O_3$, as shown in Fig. 1(c)[4-6], and subsequently argued for organic Mott-based superconducting κ-(BEDT-TTF)$_2$Cu[N(CN)$_2$]Cl[7-9], as displayed in Fig. 1(d). Above the Mott endpoint (MEP), where the MIT line ceases at the critical temperature, an SCF phase was experimentally observed. The nature of the SCF phase near and above the MEP has been intensively investigated so far only using a scaling theorem based on electrical conductivity. An SCF of carrier-tuned Cr-$V_2O_3$ behaves like a classical fluid, similar to a G-L SCF exhibiting a three-dimensional (3D) classical Ising criticality[4-6]. On the other hand, volume-controlled Mott-based organic conductors have recently been reported to exhibit a distinct quantum-critical behavior with unconventional critical exponents[7-9]. In both studies, the role of microscopic inhomogeneities, such as two-phase coexistence, percolative transport, and low dimensionality that greatly modify the critical behaviors of macroscopic scaling, remains an unanswered significant scientific question.

Alkali metal $Cs_xRb_{3-x}C_{60}$ fullerides with expanded cell volumes, having recently joined a club of Mott materials, exhibit an MIT from a paramagnetic Mott-insulator to a paramagnetic metal without symmetry breaking and an unconventional superconductivity (SC) with a SC $T_c$-dome[10-13]. Considering the unique features of unchanged carrier number of N=3 per $C_{60}$ and 3D nature, they are expected to provide an excellent research platform for elucidating the intrinsic quantum nature of MIT in Mott materials. However, unlike Cr-$V_2O_3$ and κ-(BEDT-TTF)$_2$Cu[N(CN)$_2$]Cl, physical properties were studied by only magnetic probes without electrical transport data, and neither an SCF nor an MEP has been observed, as shown in Fig. 1(e)[13].

In the present work, we conduct the first systematic study of electrical conductivities (σ) under both ambient and high pressure (P) by focusing on $Cs_xRb_{3-x}C_{60}$ fullerides near the Mott-insulator boundary, in addition to extended static magnetic susceptibilities (χ), with a strong motivation to search for the missing SCF phase and MEP. Electrical transport probes exhibit a new MIT line, distinct from that observed by magnetic probes. The MEP's existence is unambiguously confirmed by the single point where the two distinct MIT lines meet. The unambiguous two sets of experimental data (χ and σ), which are not available for other Mott materials (see Fig. S7), persuade us to make novel dual analyses on the nature of the SCF of Mott materials. Our analyses of χ, based on a model Hamiltonian, yield an unambiguous hypothesis-free conclusion that the SCF phase of $Cs_xRb_{3-x}C_{60}$ fullerides is an M-I coexisting SCF, exhibiting a novel MIT line (Widom line[14] in SCF) with an I and M ratio $x_I/x_M$ = 1. This can be compared with the results available from the scaling analysis based on σ, with assuming a uniform single phase, which yields a Widom line exhibiting a critical exponent in the quantum regime. These remarkable conclusions of the SCF phase in $Cs_xRb_{3-x}C_{60}$ fullerides (Fig. 2(a)) not only provide deep new insights into the SCFs but also share the key features with the pseudo-gap (PSG) phase commonly discussed in other high $T_c$ Mott materials, such as cuprates (Fig. 2(b))[15] and two-dimensional (2D) moiré materials[16], by complementing the state-of-the-art theoretical calculations[17-22].

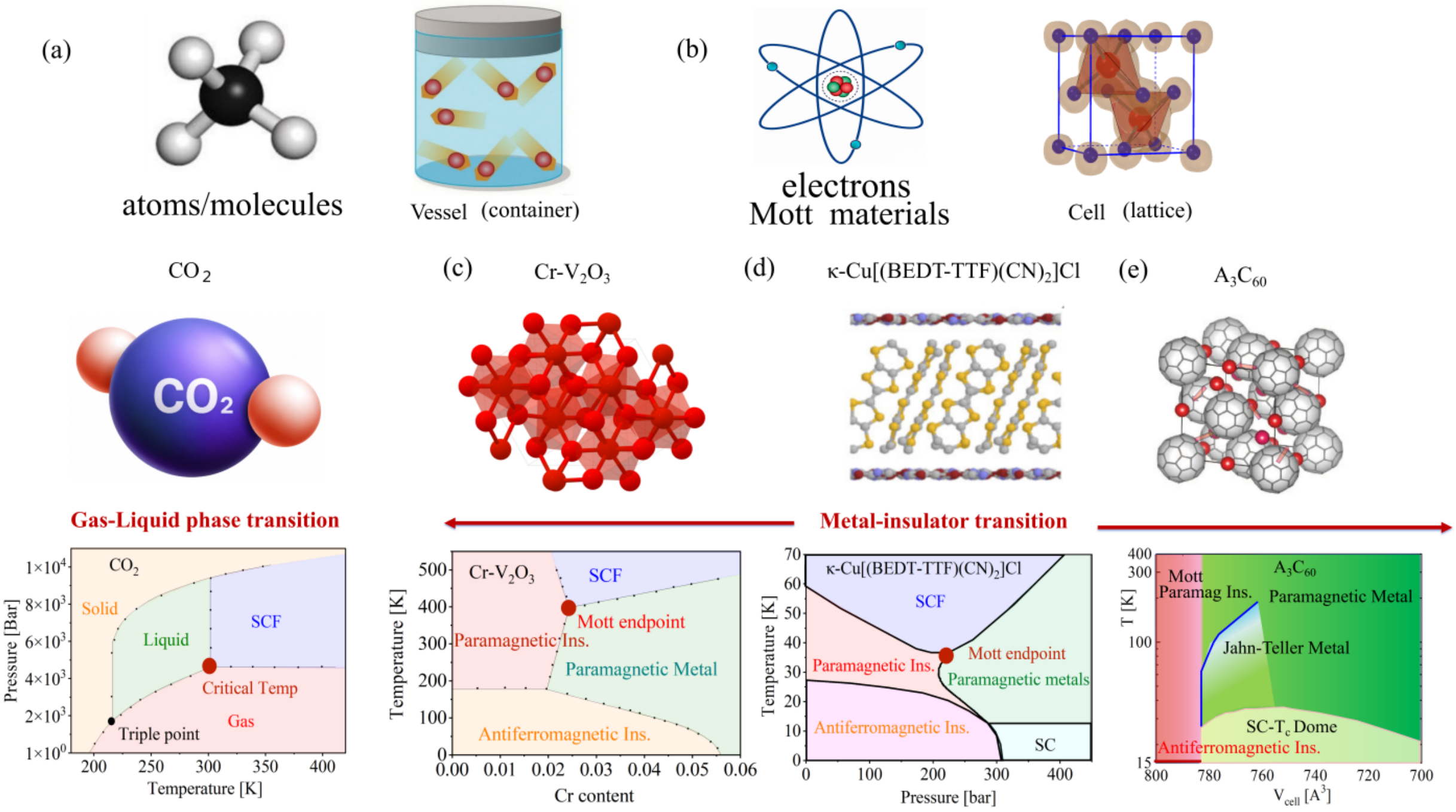

**Fig. 1 Phase transition without symmetry breaking and supercritical fluid phase (SCF).|** (a) Phase transition between G and L phases in elemental atoms and molecules in a container without symmetry breaking and a supercritical fluid (SCF) phase emergent above the critical temperature (a red point). (b) Quantum electrons, in a crystal cell. In Mott materials ((c) Cr-$V_2O_3$ and (d) κ-(BEDT-TTF)$_2$Cu[N(CN)$_2$]Cl), a similar phase diagram exhibiting an SCF phase is reported. The SCF phase originates from the Mott endpoint (MEP, red-colored point). (e) $Cs_xRb_{3-x}C_{60}$ fullerides, Mott unconventional superconductors, exhibit an MIT (blue line), but neither MEP nor SCF was reported.

## Supercritical fluid (SCF) phase

The phase diagram of $Cs_xRb_{3-x}C_{60}$ fullerides[10-13] has been carefully re-examined based on the newly measured σ's in addition to χ's as a function of T and V, as shown in Fig. 2(a). We plotted various experimental points exhibiting MIT, based on the σ(T, V)'s for $Cs_xRb_{3-x}C_{60}$ fullerides at ambient and under high-pressure conditions. A new MIT line (red color) based on σ's was vastly different from a blue MIT line based on χ's. These two distinct MIT lines originate from a common point in the T-V phase diagram, which is attributed to an MEP (the red-filled circle in Fig. 2(a)). An SCF phase in $Cs_xRb_{3-x}C_{60}$ fullerides becomes emergent as a zone (pale blue) above the MEP. The newly discovered SCF phase by our present experiments is located in the insulating zone previously indicated by magnetic probes and X-ray structural analyses[13].

The difference between the two MIT lines (blue and red) becomes considerably larger with $V_{cell}$, spanning from a large lattice in $Cs_{2.65}Rb_{0.35}C_{60}$ to a small one in $Cs_2Rb_1C_{60}$, before fully transitioning into a good Fermi liquid metal of $Cs_1Rb_2C_{60}$ (green) and $Rb_3C_{60}$ (blue) as displayed by the three different color zones of Fig. 2(a). In $Cs_2Rb_1C_{60}$, the observed difference in the MIT onset temperature between σ and χ is more than 100 K (orange color). This is far beyond a general difference of less than 1 K and even smaller in conventional materials. One may attribute the difference in the onset temperatures between the two observation probes to the impurities or stoichiometric deviations. However, such speculations can be safely ruled out, because purely sharp superconducting transitions were observed for $Cs_xRb_{3-x}C_{60}$ fullerides with identical $T_c$'s between σ and χ within the conventional experimental errors (see Fig. S1).

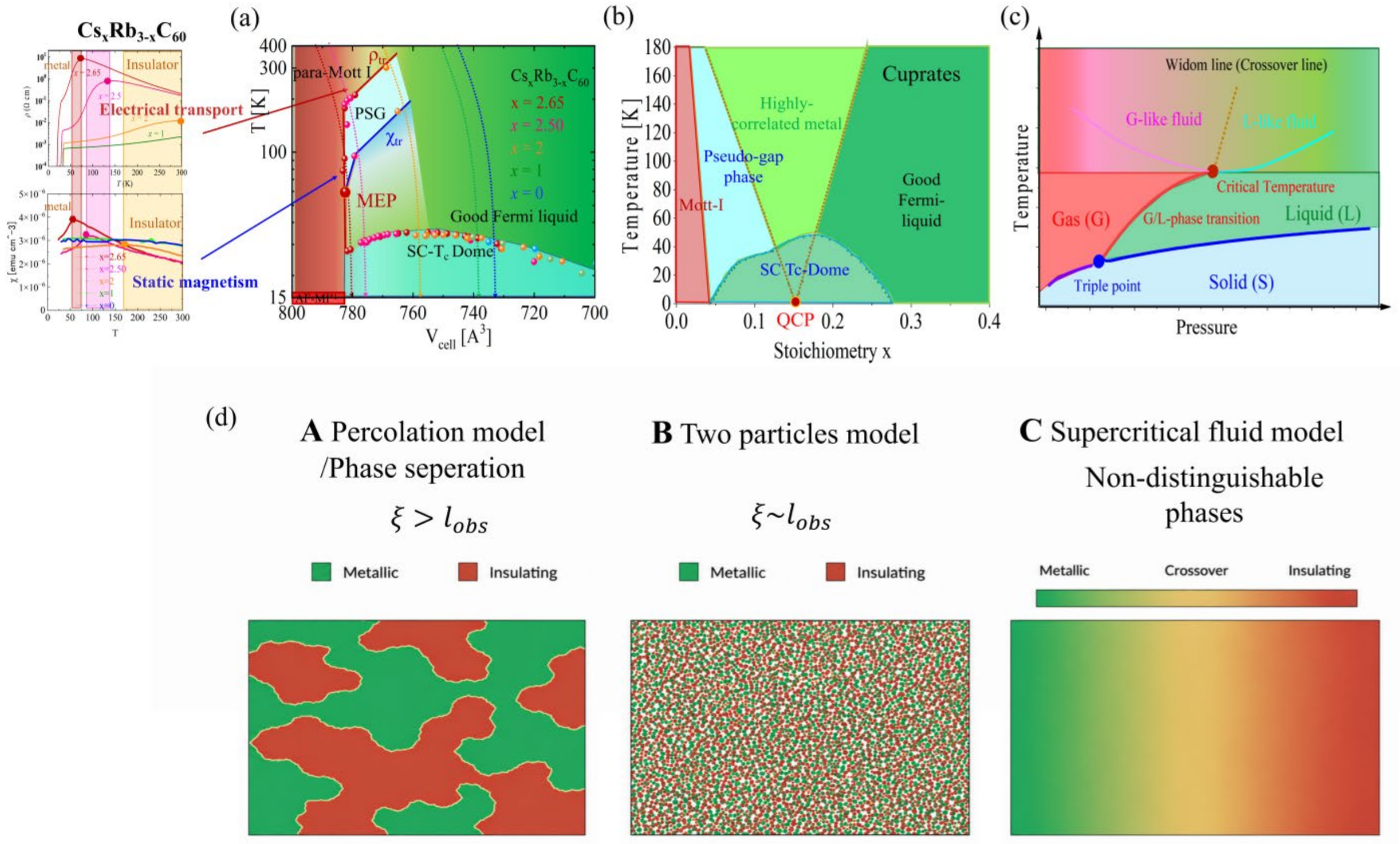

**Fig. 2 T-V(T, P) phase diagram of $Cs_xRb_{3-x}C_{60}$ fullerides displaying an M-I coexisting SCF phase inferred from electrical resistivities and static magnetic susceptibilities.**| (a) The onset temperatures of the MIT observed in ρ under ambient and high pressure of $Cs_xRb_{3-x}C_{60}$ fullerides give a different MIT line (red) from the one (blue) drawn based on χ. The two lines originate from the single common point, i.e., the so-called Mott endpoint (MEP). The same SC $T_c$-dome is observed, within experimental accuracy, in both ρ and χ. (b) Phase diagram of carrier-tuned high-$T_c$ cuprates generally reported in the literature[15], where a PSG phase observed at finite T is frequently discussed in connection with the QCP in the T=0 limit. (c) A general phase diagram of SCF observed in Ar as a function of P, where a Widom line exhibiting a crossover line from G-like to L-like fluid is displayed. (d) Three typical models of an M-I coexisting phase with different correlation lengths ξ: (A) $\xi > l_{obs}$ (B) $\xi \simeq l_{obs}$ (C) $\xi < l_{obs}$, where $l_{obs}$ denotes the length of the experimental observation limit.

Since theoretical quantitative evaluation is possible for magnetism different from the scaling on electrical conductivity, we apply a two-particle model for the dual properties of χ in SCFs, similar to a G/L system of elemental atoms[14], $\chi^{obs} = x_M\chi_{Pauli} + x_I\chi_{CW}$ ($x_M + x_I$=1). Here $\chi_{Pauli}$ and $\chi_{CW}$ are the Pauli paramagnetic susceptibility term of itinerant electrons and the Curie-Weiss term of localized electrons, respectively. The Curie-Weiss term of $\chi_{CW} = (1/3)g_e^2\mu_B^2S(S+1)/(k_B(T-\Theta))$, where S=1/2 according to the electronic configuration in the Mott ground state[17-23], $g_e$=2.0023 is the g-factor of a free electron, $\mu_B$ is the Bohr magneton, and $k_B$ the Boltzmann constant. $\Theta$ defines the Curie-Weiss temperature, which originates from antiferromagnetic super-exchange interactions with site (i, j) and orbital (a, b = x, y, z) dependent couplings $J_{ij,ab} \simeq 8t_{ij,ba}\, t_{ij,ab} / (U+U')$. Here, U and U' are the intra(inter)-orbital Hubbard Coulomb repulsion and $t_{ij,ab}$ are the transfer integral elements for two $C_{60}$ molecules at positions $\mathbf{R}_i$ and $\mathbf{R}_j$. Since we are interested in the regime above the MEP, where the dynamic Jahn-Teller (JT) effect occurs, we assume that each spin with S=1/2 is equally distributed among the three $t_{1u}$ orbitals of each molecule, with $\mathbf{S}_{ix} = \mathbf{S}_{iy} = \mathbf{S}_{iz} = \mathbf{S}_i /3$. After performing the sums over orbitals, we obtain the effective spin Hamiltonian $H_J = ½ \Sigma_{i\neq j} J_{ij}\, \mathbf{S}_i \cdot \mathbf{S}_j$, with $J_{ij} = \Sigma_{a,b} J_{ij,ab} /9$. This enables us to evaluate $\Theta \simeq$ {-119.4, -98.6, -89.4, -71.0, -58.2} K for $V_{cell}$ = {722, 750, 762, 784, and 804} Å$^3$. See Methods for additional details. We calculated $\chi^{obs}$ as a function of $V_{cell}$ using an interpolation over Θ's. We remark that we do not associate the calculated Θ's with the experimentally observed Néel temperature, since the latter is below 2 K due to the geometrical frustration of face-centered cubic symmetry of $Cs_xRb_{3-x}C_{60}$ fullerides[13,23]. The Pauli magnetic susceptibility term takes the form $\chi_{Pauli} = \mu_B^2 D_{EF}/(1-(U+U')D_{EF})$, including Stoner's enhancement factor. In the present calculations, we extended our nearly T-independent experimental data of $\chi_{Pauli}(T, V_{cell})$ of the metallic $CsRb_2C_{60}$ fulleride to other fullerides.

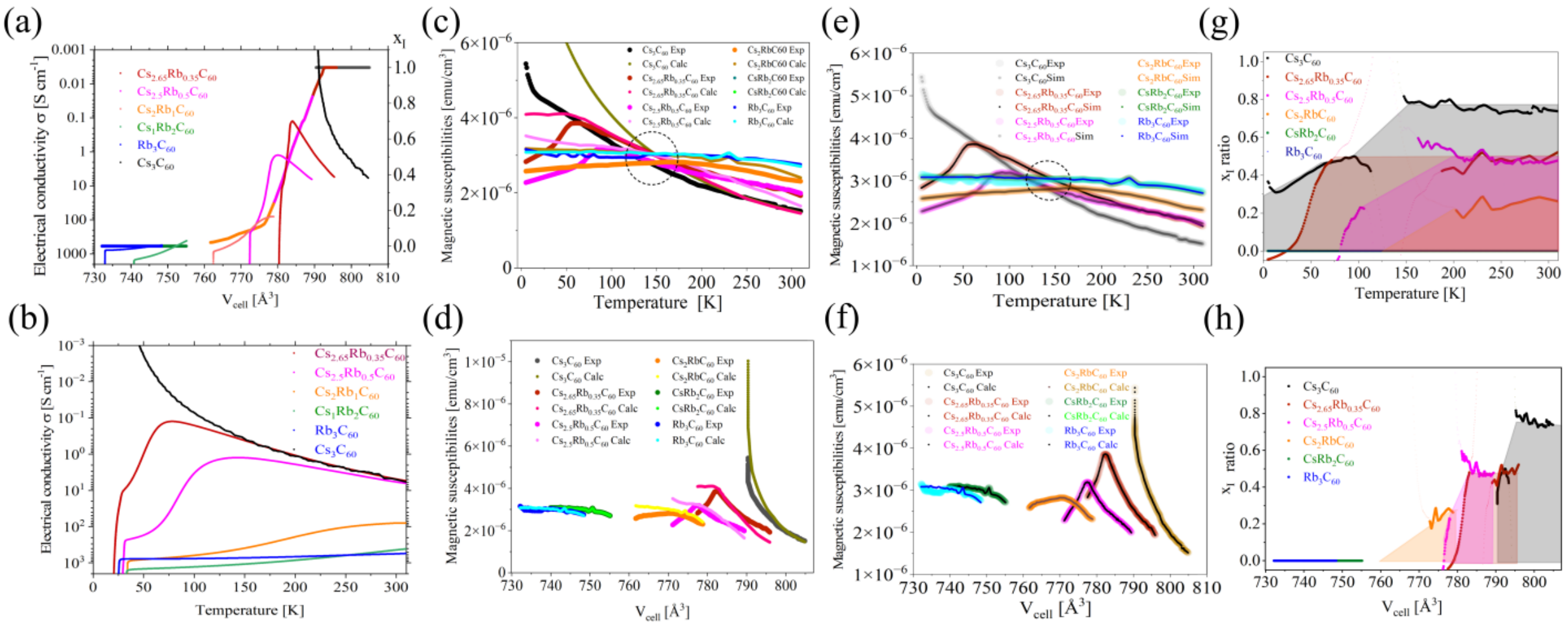

**Fig. 3 The experimental σ(T, V) and χ(T, V) values for various stoichiometric $Cs_xRb_{3-x}C_{60}$ fullerides under ambient pressure.**| Three $Cs_{2.65}Rb_{0.35}C_{60}$ (red), $Cs_{2.5}Rb_{0.5}C_{60}$ (pink), and $Cs_2Rb_1C_{60}$ (orange) exhibit the clear MIT's. (a,b) The σ values of various stoichiometric $Cs_xRb_{3-x}C_{60}$ fullerides under ambient P as a function of $V_{cell}$ and T. The values χ and σ of $Cs_3C_{60}$ were taken from the literatures[13,24], respectively. (c-f) The comparison of χ between experimental (pale wide lines) and calculated (dark narrow lines) data. The analyses were performed as $x = x_I\chi_{CW} + x_M\chi_{Pauli}$ with $x_I + x_M = 1$, where $x_I$ and $x_M$ are the ratios of I and M, and $\chi_{CW}$ and $\chi_{Pauli}$ are the localized Curie-Weiss and itinerant Pauli magnetic susceptibilities. The Curie-Weiss temperatures (Θ) as a function of $V_{cell}$ were calculated based on the model Hamiltonian. The details are described in the main text and Methods. (c,d): Comparison between experimental data and calculated ones, when $x_I$ as a function of $V_{cell}$ is given according to a linear-connection line of the MIT points observed in σ. (g,h): Evaluated $x_I$ and $x_M$, when the calculated

$\chi^{calc} = x_I\chi_{CW} + x_M\chi_{Pauli}$ with $x_I + x_M = 1$ provides the best fit to the experimental $\chi^{obs}$ using $x_I$ ($x_M$) as the free parameter. The important result of the equilibrated Widom line with $x_I/x_M = 1$ emerges without any hypotheses, providing an essential experimental evidence that the SCF in $Cs_xRb_{3-x}C_{60}$ fullerides includes an equilibrium M-I coexisting SCF phase by exhibiting the maximum missing entropy as described in the text.

First, we performed our analyses using the $x_I$ and $x_M$, by approximating the linear MIT line as shown in Fig. 3(a). We theoretically calculated $\chi^{obs}$ values as explained earlier and compared them with our experimental data in Fig. 3(c) and (d). Although our calculations describe the experimental data fairly well, a deviation remains. Next, we evaluated the experimental $\chi^{obs}$ by treating $x_I$ (= 1 - $x_M$) as a variable with the constraint of $x_I + x_M = 1$. The calculations showed an excellent agreement with our experimental data of $\chi$, as shown in Fig. 3(e) and (f). The resulting values of $x_I$ are provided in Fig. 3 (g)-(h) as a function of T and $V_{cell}$.

We identify the following significant features in our analysis. (1) All experimental $\chi^{obs}$ data intersect at nearly the same temperatures, as shown by a circle in Fig. 3(c) and (e). The relationship in $\chi^{obs}$ yields the intersection temperature $T^* = \mu_B^2/(k_B V_{cell}) + \Theta$ using the S=1/2 and $g_e$=2.0, when $\chi_{Pauli} = \chi_{CW}$. The calculated T* values using the {$V_{cell}$, $\Theta$} obtained from the model Hamiltonian are T*={146K for $Cs_3C_{60}$, 138 K for $Cs_{2.65}Rb_{0.35}C_{60}$, 118 K for $Cs_2RbC_{60}$}, providing a reasonable explanation for the experimental data. (2) The $\chi^{obs}$ values at 300 K gradually increase from insulating $Cs_3C_{60}$ to metallic $Rb_3C_{60}$. Although the MIT temperatures in our new experiments are the same as those in the previous report[13], the $\chi$ intensities differ (see Fig. S3-1) and are essential for providing quantitative interpretation. (3) The magnetic susceptibilities of $Cs_{2.65}Rb_{0.35}C_{60}$, $Cs_{2.5}Rb_{0.5}C_{60}$, and $Cs_2RbC_{60}$ are smaller than $\chi_{Pauli}$ values of $CsRb_2C_{60}$ and $Rb_3C_{60}$ at low temperatures before entering the SC states. This indicates that the antiferromagnetic interactions originating from U and U' suppress the magnetic response.

The results described above are indeed profound. As we can discern from these figures, the crossover Widom line[14] possessing $x_I/x_M = 1$ arises from the quantitative analyses of magnetic susceptibility data, indicating the maximum of two-phase mixing entropy $S_{mix}$ ($x_I/x_M = 1$ is the solution of $dS_{mix}/dx_I = 0$, where $S_{mix} = x_I\ln(x_I) + x_M\ln(x_M)$ with $x_I + x_M = 1$, which indicates a thermodynamical equilibrium). Our results based on $\chi$ constitute unambiguous experimental evidence, free of any hypothetical bias, as a new distinct quantitative approach to the scaling of electrical conductivity, and firmly indicate that both phases of M and I coexist in the SCF under thermodynamic equilibrium, not as a uniform single phase. The newly assessed Widom line ($x_I/x_M = 1$) is crucial for supporting a two-particle model B, since it rules out the possibility of neither a separated nor percolation-like phase segregation model A nor a non-distinguishable uniform phase model C in Fig. 2(d).

To be compared with the present quantitative analysis of the magnetic susceptibilities based on a two-particle model, we have separately performed a finite temperature scaling analysis based on electrical resistivities of $Cs_{2.65}Rb_{0.35}C_{60}$ measured under various P's (Fig. S4) in a similar fashion to the previous literatures[8,25], as shown in Fig. 4(a) (the details are given in SI.9). The results of scaling of $\rho(T,V-V_c)/\rho(T_0,0) = F[T/|C(V-V_0)|^{z\nu}]$ in temperature region of $T>T_0$, where ν and z are the correlation length and dynamic critical exponents, $T_0$ is the finite critical temperature, C is an arbitrary constant, and F is a scaling function. Our analyses provide the quantum components $z\nu = 0.6$, which differs from the classical three-dimensional Ising-type critical exponent of $z\nu = 1.2$[7,26] and is similar to those reported for the quantum critical MIT phenomenon in Mott materials[8,9,25]. Our comparative analysis of two different analyses has successfully provided the intrinsic picture of the SCF phase that has not been deduced solely from the scaling analysis.

## Gibbs phase rule in T-P(T,V) phase diagram

The Gibbs phase rule[27] (F=C-D+2, where F is the freedom of a system, C the number of components, and D the number of phases) is one of the most fundamental principles of a phase diagram. In one-component materials (C=1, F=2, and then D=1), an M-I coexisting phase can only be permitted as a line. Consequently, a dual-nature phase without such a distinguishable two-phase boundary above the critical temperature has been conceptually explained by SCF[1-3]. $Cs_xRb_{3-x}C_{60}$ fullerides exhibit an SCF-like phase above MEP. X-ray structural[13] and NMR analyses[28] show that the materials are homogeneous and do not indicate phase separation within the resolution of experimental observations. Both magnetic susceptibility and electrical transport studies show a smooth and continuous evolution as functions of T and P, consistent with an unidentified single crossover phase transition. Our scaling analyses of electrical transport data, assuming a single homogeneous SCF phase above the MEP, suggested an unconventional quantum critical crossover phenomenon characterized by a quantum critical exponent $z\nu = 0.60$, similar to those of organic conductors[8,9]. Our new quantitative analysis of magnetic properties, on the other hand, evidenced the dual nature of M-I coexisting phases. In the next section, we extend our interpretation by adding orbital freedom as the generalized Gibbs phase rule, thereby permitting a two-phase coexistence under thermodynamic equilibrium[29,30].

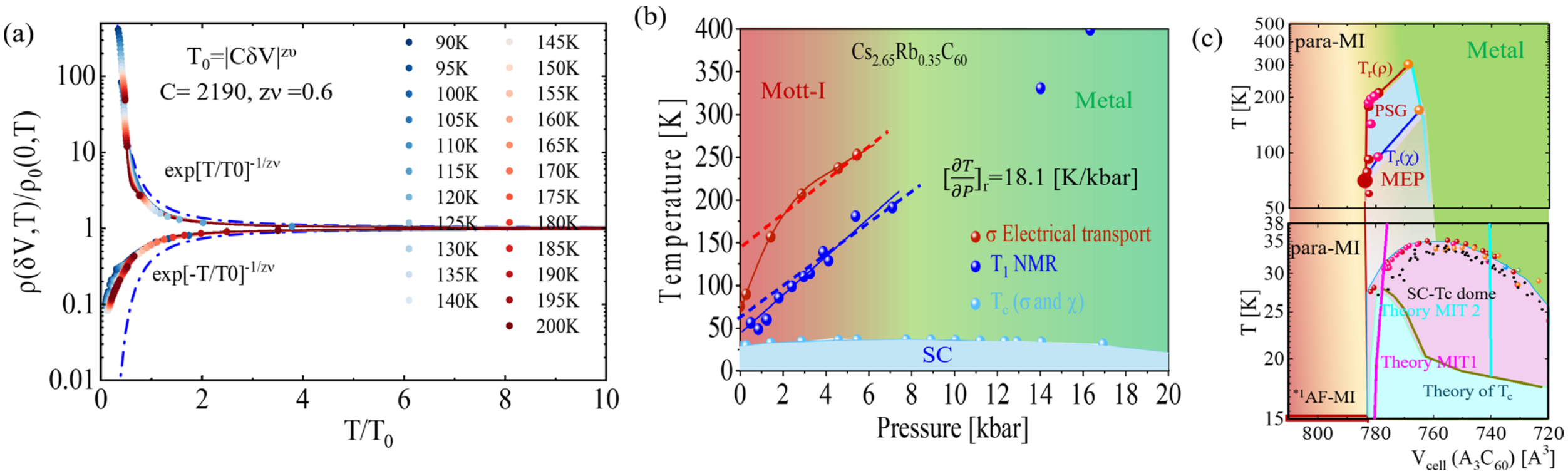

**Fig. 4 Scaling based on electrical transports, Clapeyron Clausius Law, and T-V(T, P) phase diagram with comparison to DMFT theoretical calculations.|** (a) Scaling of $\rho(V-V_0,T)/\rho_0(T,0) = F[T/(C(V-V_0))^{z\nu}]$ based on electrical transport data in $Cs_{2.65}Rb_{0.35}C_{60}$ under various pressure (SI.4 and SI.9). (b) The slope of the MIT ($\partial T/\partial P$) exhibits a good agreement between electrical transport (our work) and NMR relaxation times[28] at temperatures under the dynamic Jahn-Teller effect as described in the text. The MIT slope can be assessed from the volume ($\Delta V_{cell}$) and the entropy change ($\Delta S$) at

MIT, using Clapeyron-Clausius's law, as explained in the text and SI.6. (c) T-V phase diagram with SC-$T_c$ dome. The two lines (purple and light blue), calculated by the first-principles calculations reported in literature[19], can be assigned to the MIT line (red) at low T's and the border line (light-blue) between M-I coexisting SCF and normal metal, as explained in the text.

## Interpretations shared with recent theories

Sophisticated first-principles theoretical calculations[19] predicted two MIT lines (purple and sky-blue lines) in Fig. 4(c), where the middle region of the two lines becomes insulating or metallic depending on the initial conditions of the calculations. The previous experimental T-V phase diagram, on the other hand, exhibited a single blue MIT line[13] with neither the MEP nor the SCF phase, in contrast to our new T-V phase diagram, which exhibits two MIT lines (blue and red). A new alternative interpretation can presently be given for the theoretical calculations. One theoretical MIT line (purple) corresponds to the experimental MIT line (red), whereas the other theoretical line (sky blue) is ascribed to the boundary between the M-I coexisting SCF and a good Fermi-liquid metallic phase, as indicated in Fig. 4(c).

Another critical discussion concerns the slope of the MIT line. The MIT line observed in our electrical transport measurements for $Cs_{2.65}Rb_{0.35}C_{60}$ under various pressures (see Fig. S4) is shown in Fig. 4(b), together with the previous NMR report[28]. The slope of MIT under thermodynamic equilibrium can be understood in terms of the Clapeyron-Clausius law of $(\partial P/\partial T)_{MIT} = \Delta S/\Delta V$ (see the details in SI.6), where ΔS and ΔV are the changes of entropy and volume at MIT. Based on the σ(T, P) of $Cs_{2.65}Rb_{0.35}C_{60}$, the MIT occurs at around P = 0.6 - 0.8 GPa (see Fig. S4), and its ΔV is reported to be c.a. -5Å$^3$/$V_{cell}$ (= -3.01cm$^3$/$C_{60}$-mol)[13]. The P evolution of $(\partial P/\partial T)_{MIT}$ = 65 ~ 15 K/kbar (see Fig. S6) gives $\Delta S = \Delta V/(\partial T/\partial P)_{MIT} \times 100/4.184$, which increases from 1.1 to 4.8 cal/(K $C_{60}$-mol) with a plateau of ΔS= 2.5 cal/(K $C_{60}$-mol) (Fig. S6(c)). In theory, Δ$S$ at MIT can be evaluated using Boltzmann's entropy formula to be $\Delta S = N_A k_B \ln 3$ = 2.2 [cal/(K $C_{60}$-mol)] (see Methods and SI.6), where $N_A$ is Avogadro's number. In the dynamic JT effect at higher temperatures, where the time scale of the electron transfer among the JT states is faster than the observation time of the electronic states, three degenerate $t_{1u}$ orbitals are treated as equivalent. However, the orbital freedom is gradually lost at lower temperatures upon the change in the JT effect from dynamic to static, and the situation is modified. A large slope $(\partial T/\partial P)_{MIT}$ at lower temperatures than $T_0$ in the T-P phase diagram reflects such a situation.

Finally, we discuss the SC-$T_c$ dome. The experiments on both σ and χ exhibit an SC-$T_c$ dome, similar to those in other Mott-based high-$T_c$ superconductors, such as Cu oxides[15,31], Fe pnictides[32], and nickelates[33]. The electronic phase instability of these high-$T_c$ materials is frequently discussed in terms of a pseudo-gap (PSG) phase[34-36]. Initially, the PSG phase attracted intense interest among scientists as a preformed boson state leading to a coherent Cooper-paired superconducting state (Bose-Einstein condensation BEC mechanism), as implicated by Anderson's valence bond theory[37]. However, because multiple origins of the PSG have recently been reported, the scientific understanding of PSG in Mott-based high-$T_c$ materials has become more complex[38]. From the viewpoint of similarity among these phase diagrams reported so far, the M-I coexisting SCF phase found at finite temperatures in $Cs_xRb_{3-x}C_{60}$ fullerides is similar in position to the PSG phase in other high $T_c$ Mott materials, as shown in Fig. 4(c). The PSG observed at finite temperature in Mott-based material is often discussed in terms of the quantum critical point (QCP) associated with the spin degrees of freedom at T=0 limit. In the case of $C_{60}$ fullerides, the PSG appears from the freedom of orbitals at finite temperatures. These contrasts in scientific views of understanding may prompt further intriguing scientific debates.

A new M-I coexisting SCF without symmetry breaking in $Cs_xRb_{3-x}C_{60}$ was firmly assessed. In a single-component SCF in general, the coexisting two phases are not permitted, and no steady equilibrium G/L (I/M in the case of MIT) ratio is defined. Many experiments, however, detect G- versus L-like coexistent phases using neutron scattering, optical spectroscopy, and other recently emerging and rapidly advancing methods. Our new quantitative analyses of $\chi$, based on a model Hamiltonian, revealed an M-I coexisting SCF with an $x_M/x_I=1$ crossover Widom line, indicative of maximum mixing entropy between the two particles without any additional hypotheses. On the other hand, the scaling analysis based on $\sigma$, assuming a single uniform crossover phase, yielded a quantum critical exponent $z\nu$ = 0.60 in the quantum regime. Despite microscopic two-phase coexistence, electrical transport remains governed by quantum critical fluctuations. Our new dual-probe comparative analyses for the same Mott material provide the first significant and essential experimental evidence for the SCF phase of strongly correlated electrons in the Mott materials.

## Methods

### Materials preparation

$Cs_xRb_{3-x}C_{60}$ fullerides were prepared by mixing Cs and Rb with $C_{60}$ powder in a Ta cell and heated at approximately 670 K with repeated intermittent shaking after heating under accurate temperature control, according to the method described previously. The quality of the materials was confirmed by X-ray powder diffractions with Rietveld refinement.

### Electrical transport and static magnetic measurements

The prepared powder materials of $Cs_xRb_{3-x}C_{60}$ fullerides were carefully ground to increase homogeneity of the materials and pelletized into a round shape of 2mm in diameter under high pressure. Contact electrodes of four-probe measurements were made by Au paste, and mounted on a home-made high-pressure cell at the Advanced Institute of Materials Research (AIMR) of Tohoku University, Japan, and a modified high-pressure cell of Quantum Design at Beijing Academy of Quantum Information Sciences (BAQIS), China. Electrical transport was measured using PPMS (Quantum Design).

Static magnetic susceptibilities of materials were measured using powder samples by MPMS (QD). Materials of approximately 50 mg were wrapped in a plastic film and mounted in the middle position of a plastic straw by fabricating folded parts beneath and above the sample to be placed in a symmetrical position. Superconducting diamagnetic susceptibility was measured by subtracting the two measurements between B=0 and 10 Oe by correcting the remaining magnetic field. Pauli magnetic susceptibilities were measured by applying high magnetic field and subtracting the two measurement data between 2 and 5 Tesla.

**Theoretical Analysis**

To describe the M-I coexisting SCF phase, we take into account the $t_{1u}$-orbital degrees of freedom, in addition to the two intensive thermodynamic variables T and P. The electronic ground states of a $C_{60}$ molecule with N electrons |G: N> are determined by the delicate energetic balance dictating intra- and inter-orbital electron correlations[17-22]. Based on the previous sophisticated DFMT calculations[17-19], we consider a model Hamiltonian: $H = H_T + H_U + H_{U'} + H_P + H_J = \Sigma_{i\neq j,a,b,\sigma} C^\dagger_{i,a,\sigma} t_{ij,ab} C_{j,b,\sigma} + U \Sigma_{i,a} n_{ia\uparrow} n_{ia\downarrow} + (U'\text{-}J/2)\Sigma_{i,a<b} n_{ia} n_{ib} + J \Sigma_{i,a\neq b} C^\dagger_{i,a,\uparrow} C^\dagger_{i,a,\downarrow} C_{i,b,\downarrow} C_{i,b,\uparrow} - 2J \Sigma_{i,a<b} \mathbf{S}_{ia} \cdot \mathbf{S}_{ib}$. $H_T$, $H_U$, $H_{U'}$, $H_P$, and $H_J$ define the transfer integrals, the intra-orbital and inter-orbital Hubbard Coulomb repulsion, the pair hopping interaction, and the Hund's coupling, respectively. $C^\dagger$ and C denote the creation and annihilation operators of electrons, $t_{ij,ab}$ denotes the transfer integral elements with $\mathbf{R}_j - \mathbf{R}_i$ being the distance vector of two $C_{60}$ molecules at positions $\mathbf{R}_j$ and $\mathbf{R}_i$, and a, b, σ are the indices of the three $t_{1u}$ (a, b = x, y, z) orbitals and up/down spins (σ = ↑, ↓), respectively. Here, we introduced the orbital and spin resolved number operator $n_{i,a,\sigma} = C^\dagger_{i,a,\sigma} C_{i,a,\sigma}$, the spin-summed orbital resolved number operator $n_{i,a} = \Sigma_\sigma C^\dagger_{i,a,\sigma} C_{i,a,\sigma}$, and the orbital resolved spin operator $\mathbf{S}_{ia} = ½ \Sigma_{\alpha,\beta} C^\dagger_{i,a,\alpha} \boldsymbol{\sigma}_{\alpha\beta} C_{i,a,\beta}$ with the vector of spin Pauli matrices $\boldsymbol{\sigma}$ ($\sigma_x$, $\sigma_y$, $\sigma_z$). The interaction coefficients U, U', and J are related through U'=U-2J, which are obtained after integrating out the phonons and inter-molecular Coulomb interaction $H_V = V \Sigma_{i\neq j} n_i n_j$, where $n_i = \Sigma_a n_{i,a}$. When we neglect the phonon contributions, first-principles calculations provide U ~ 1 eV, J ~ 0.05 eV, and a band width W ~ 0.5 eV[17]. However, it was later shown that the presence of Jahn-Teller e-ph interactions leads to qualitatively crucial modifications. After accounting for the e-ph coupling and the inter-site Coulomb term, the most reliable renormalized values are reported as U = 558 and J = -17 meV[18,19].

We employed the above model Hamiltonian to understand the phase diagram in the vicinity of the Mott insulator phase, where the electronic spectral weight is substantially reduced, and the physics is dictated by the ground state of isolated molecules. By studying the onsite part of the Hamiltonian, i.e., $H_{onsite} = H_U + H_{U'} + H_P + H_J$, we find that the ground state is obtained as a linear combination of the two spin-degenerate states |G: N=3> = {(|↑↓,↑,e>+|e,↑,↑↓>)/$\sqrt{2}$, (|↑↓,↓,e>+|e,↓,↑↓>)/$\sqrt{2}$}. This reflects the emergence of a single spin with S=1/2, along with an empty state labelled as "e" and a pair of electrons. The Hund's rule is violated due to the unusual energy hierarchy U < U'. The two paired electrons in a given molecule appear at the same orbital, thus giving rise to a spin-zero bosonic excitation characterized by a short coherence length. The unpaired electron with S=1/2 governs the magnetic properties of each molecule, while the paired electrons are nonmagnetic.

The ground state and the formation of superconductivity are governed by two additional factors, i.e., the Jahn-Teller effect and the presence of pairwise electron transfer. When the static Jahn-Teller effect occurs, the $t_{1u}$ orbitals are no longer degenerate, and the unpaired electron occupies a fixed orbital. In contrast, the $t_{1u}$ orbitals become degenerate under the dynamic Jahn-Teller effect. As a consequence, experimental measurements at frequencies below those of the Jahn-Teller phonons average over all three $t_{1u}$ orbitals. Hence, when the dynamic Jahn-Teller effect sets in, the two spin S=1/2 states introduced above are defined as an arbitrary choice of all equivalent orbitals for the unpaired electron. In contrast, at low temperatures, the orbital structure of the hopping part of the Hamiltonian establishes a preferred orbital character. Another important aspect concerns the presence of pair-wise electron transfer. As described above, each spin state is a superposition of two states obtained by mixing the empty and paired states. Notably, such mixing plays a vital role in the so-called Suhl-Kondo mechanism and has been theoretically shown to be crucial for the emergence of unconventional s-wave superconductivity[17-22,39,40]. All these key aspects provide a valuable framework for the discussions and interpretations described in the main text.

We consider the effects of the so far discarded hopping term $H_T$. In the Mott regime, we treat the electron transfer integrals in a perturbative manner, since these also scale with the electron weight and are suppressed in the Mott insulator phase. We focus on the case $J < 0$ and $U' > U$, and consider pairs of $C_{60}$ molecules, located at positions $\mathbf{R}_j$ and $\mathbf{R}_i$, with each one having the spin-degenerate ground state |G: N=3> mentioned above. The respective joint ground state of two decoupled molecules is $|G_{Decoupled}\rangle = |G: N=3\rangle_i \otimes |G: N=3\rangle_j$. The two $C_{60}$ molecules are solely coupled via electron transfer, since inter-molecular Coulomb interactions have already been accounted for in the renormalized onsite interaction coefficients U, U', and J as described earlier. The operation $H_T|G_{Decoupled}\rangle$ generates a subspace of states, which have two electrons in one $C_{60}$ molecule and four electrons in the other. The subspace states include various non-magnetic spin-singlet states, which are the ones to primarily contribute to virtual processes mediated by the electron transfer. These spin-singlets are constructed by electron pairs residing in the same or different orbitals.

By virtue of the unusual energy hierarchy U'>U as well as $|J| \ll U$ and U', the energies of these virtual excited states are quite close to each other. For exploiting this property, we employ the Brillouin-Wigner perturbation theory and arrive at an effective antiferromagnetic super-exchange coupling for the two spins between two molecules. The coupling is described as the following form $H_J = \frac{1}{2} \Sigma_{i\neq j,a,b} J_{ij,ab} \mathbf{S}_{ia} \cdot \mathbf{S}_{jb}$, where $J_{ij,ab} \simeq 8t_{ij,ba} t_{ij,ab} / (U+U')$, and fully determines the nature of the Mott phase obtained by accounting for virtual transitions to the excited states of two decoupled molecules. The above simplified and compact form for the super-exchange interaction is obtained by assuming that all the excited states lie roughly at the same energy, that is, (U+U')/2 higher than the ground state energy $E_G = 6U'+4J$ of two decoupled molecules. Detailed information concerning the state-vectors, energies, and multiplicity of all the virtual states is given in Table S1.

By minimizing the above super-exchange coupling, one can, in principle, identify the ground state of the entire crystal and infer the detailed orbital and spin structure of the molecules as well as the emergent Mott phase. Furthermore, one can in principle reach a more precise understanding of the antiferromagnetic transition and the onset temperatures of the emergent SCF phase by evaluating the critical temperature of the magnetic phase transition and investigating the super-exchange interaction at finite temperatures. However, this is a formidable task by considering the geometrical frustration widely known to dictate antiferromagnets on the face-centered cubic (fcc) lattice. Indeed, it is well established that the fcc $Cs_3C_{60}$ Mott-insulator ground state arises at a Néel temperature ~ 1.5 K, which is much smaller than the energy scale defined by the couplings $J_{ij,ab}$. From the first-principles calculations carried out earlier[19], the largest electron-transfer integral is typically of the order of ~ 40 meV. This yields a super-exchange energy scale $E_J \simeq 65$ - 70 K, being comparable to the temperature of the Mott-endpoint. While the super-exchange model may not be reliable in describing the antiferromagnetic transition, it still appears capable of describing the aspects of the SCF phases described in the text.

The potential relevance of the above super-exchange model is employed to derive a theoretical expression for the magnetic susceptibility, which is vital in the present context and enables comparisons with the experimental measurements. To study the magnetic susceptibility of the localized S=1/2 spin, we consider the multi-orbital super-exchange model $H_J = ½ \Sigma_{i\neq j,a,b} J_{ij,ab} \mathbf{S}_{ia} . \mathbf{S}_{jb} + g_e\mu_B\Sigma_{i,a} \mathbf{B}_i . \mathbf{S}_{ia}$, where we introduced a spatially varying magnetic field $\mathbf{B}_i$. The corresponding partition function of a system at T with $k_BT = 1/\beta$ is described as $Z = Tr[exp(-\beta H_J)]$, where the trace runs over the orbitals, the site indices, and the molecular spin states. Since we are mainly interested in the regime above the Mott-endpoint of the dynamic Jahn-Teller, we assume that every spin with S=1/2 is equally distributed in the three $t_{1u}$ orbitals of each molecule. This allows us to set $\mathbf{S}_{ix} = \mathbf{S}_{iy} = \mathbf{S}_{iz} = \mathbf{S}_i /3$ and perform the sums over the orbitals. This process yields the Hamiltonian $H_J = ½ \Sigma_{i\neq j} J_{ij} \mathbf{S}_i . \mathbf{S}_j + g_e\mu_B\Sigma_i \mathbf{B}_i . \mathbf{S}_i$, with $J_{ij} = \Sigma_{a,b} J_{ij,ab} / 9$.

To proceed, we adopt an auxiliary-field approach (see chapter 4 in the literature[41]), which is carried out by employing a local auxiliary field $\boldsymbol{\psi}_i$ through a Gaussian identity, which yields the partition function $Z=Z_0\int d\psi\, Tr\{exp[-\beta h(\psi)]\}$, with $h(\psi) = ½ \Sigma_{i,j} (-\mathbf{J}^{-1})_{ij} \boldsymbol{\psi}_i . \boldsymbol{\psi}_j + \Sigma_i (\boldsymbol{\psi}_i + g_e\mu_B \mathbf{B}_i) . \mathbf{S}_i$, where $\int d\psi$ runs over configurations of the auxiliary field, $\mathbf{J}$ is the corresponding coupling matrix in site space, and $Z_0$, a reference partition function. As it is customary, we now redefine the auxiliary field so that it absorbs the magnetic field, and obtain $h(\psi) = ½ \Sigma_{i,j} (-\mathbf{J}^{-1})_{ij} (\boldsymbol{\psi}_i - g_e\mu_B \mathbf{B}_i) . (\boldsymbol{\psi}_j - g_e\mu_B \mathbf{B}_j) + \Sigma_i \boldsymbol{\psi}_i . \mathbf{S}_i$. Integration of the microscopic degrees of freedom, i.e., spins, is straightforward, since this part of the problem describes a localized spin exchanged-coupled to an effective magnetic field $\boldsymbol{\psi}_i$. Since we are primarily interested in the Curie-Weiss magnetic susceptibility, we integrate out the spins and retain terms in the partition function up to quadratic order with respect to $\boldsymbol{\psi}_i$. This leads to an effective Hamiltonian $h_{eff}(\psi) = - N_c \ln(2)/\beta + ½ \Sigma_{i,j} (-\mathbf{J}^{-1})_{ij} (\boldsymbol{\psi}_i - g_e\mu_B \mathbf{B}_i) . (\boldsymbol{\psi}_j - g_e\mu_B \mathbf{B}_j) - ½ \beta S^2\Sigma_i \boldsymbol{\psi}_i . \boldsymbol{\psi}_i$, where $N_c$ is the number of unit cells.

We employ the saddle-point approximation (SPA) for the auxiliary field in order to evaluate the partition function. Within the SPA, the auxiliary field satisfies $\Sigma_j (\delta_{ij}+\beta S^2 J_{ij})\boldsymbol{\psi}_j = g_e\mu_B \mathbf{B}_i$, which leads to $h_{eff;SPA} = - N_c\ln(2)/\beta - ½ \Sigma_{i,j} \chi_{ij} \mathbf{B}_i . \mathbf{B}_j$, where we introduced the magnetic susceptibility $\chi_{ij}$ with the inverse $(\chi^{-1})_{ij} = [k_BT\delta_{ij} +(1/3)S(S+1)J_{ij}] / [(1/3)g_e^2\mu_B^2S(S+1)]$. Note that the susceptibility structure retains the standard Curie-Weiss form for an antiferromagnetic super-exchange Hamiltonian. Since the coupling $J_{ij}$ only depends on $\boldsymbol{\delta} = \mathbf{R}_j - \mathbf{R}_i$, the uniform susceptibility becomes $\chi_{CW} = (1/(3k_B))g_e^2\mu_B^2S(S+1) /(T - \Theta)$. Here $\Theta = - (1/3)S(S+1)\Sigma_{\boldsymbol{\delta}} J_{\boldsymbol{\delta}}/k_B$ generally does not coincide with $T_N$ in the case of a frustrated spin system. Since the transfer integrals are much stronger between nearest neighbors (with coordination number 12), than next nearest neighbors (with coordination number 6), we restrict to the nearest neighbor term in $J_{ij}$. By employing the results of the literature[17-19], we find the expression for the Curie-Weiss temperatures $\Theta = - (1/k_B)(4/3)(F_1^2+F_3^2+F_4^2+2F_2^2)/[(U+U')/2]$. The transfer integrals $F_{1,2,3,4}$ are in energy units and defined as in the literature[17-19]. By plugging in the values for $F_{1,2,3,4}$ obtained, we find the respective temperatures read as $\Theta \backsimeq$ {-119.4, -98.6, -89.4, -71.0, -58.2}K for $V_{cell}$ = {722, 750, 762, 784, and 804}$Å^3$. Note that the case of 722 $Å^3$ has been added for completeness since it corresponds to the $K_3C_{60}$ fulleride, which we did not experimentally study in this work. From the above results, we find that $|\Theta|$ decreases upon increasing the unit-cell volume, that is, as we move closer to the Mott phase, which is consistent with the trend observed in our experiments.

## References

1. Li, X. & Jin, Y. Thermodynamic crossovers in supercritical fluids. *Proc. Natl Acad. Sci. U.S.A.* **121** e2400313121 (2024).
2. Khmelinskii, I. & Woodcock, L. V. Supercritical fluid gaseous and liquid states: A review of experimental results. *Entropy* **22**, 437 (2020).
3. Ranieri, U. et al. Crossover from gas-like to liquid-like molecular diffusion in a simple fluid. *Nat. Commun.* **15**, 4142 (2024).
4. Limelette, P. et al. Universality and critical behavior at the Mott transition. *Science* **302**, 89–92 (2003).
5. McWhan, D. B., Rice, T. M. & Remeika, J. P. Mott transition in Cr-doped $V_2O_3$. *Phys. Rev. Lett.* **23**, 1384 (1969).
6. Lechermann, F., Bernstein, N., Mazin, I. I. & Valentí, R. Uncovering the mechanism of the impurity-selective Mott transition in paramagnetic $V_2O_3$. *Phys. Rev. Lett.* **121**, 106401 (2018).
7. Limelette, P. et al. Mott transition and transport crossovers in the organic compound κ-$(BEDT\text{-}TTF)_2Cu[N(CN)_2]Cl$. *Phys. Rev. Lett.* **91**, 016401 (2003).
8. Kagawa, F., Miyagawa, K. & Kanoda, K. Magnetic Mott criticality in a κ-type organic salt probed by NMR. *Nat. Phys.* **5**, 880–884 (2009).
9. Kanoda, K. Metal–insulator transition in κ-$(ET)_2X$ and $(DCNQI)_2M$: Two contrasting manifestations of electron correlation. *J. Phys. Soc. Jpn.* **75**, 051007 (2006).
10. Takabayashi, Y. et al. The disorder-free non-BCS superconductor $Cs_3C_{60}$ emerges from an antiferromagnetic insulator parent state. *Science* **323**, 1585–1590 (2009).
11. Ganin, A. et al. Polymorphism control of superconductivity and magnetism in $Cs_3C_{60}$ close to the Mott transition. *Nature* **466**, 221–225 (2010).
12. Ganin, A. et al. Bulk superconductivity at 38 K in a molecular system. *Nat. Mater.* **7**, 367–371 (2008).
13. Zadik, R. H. et al. Optimized unconventional superconductivity in a molecular Jahn–Teller metal. *Sci. Adv.* **1**, e1500059 (2015).
14. Simeoni, G. et al. The Widom line as the crossover between liquid-like and gas-like behaviour in supercritical fluids. Nature Phys. 6, 503-507 (2010).
15. Rybicki, D., Jurkutat, M., Reichardt, S., Kapusta, C. & Haase, J. Perspective on the phase diagram of cuprate high-temperature superconductors. *Nat. Commun.* **7**, 11413 (2016).

16. Li, T. et al. Continuous Mott transition in semiconductor moiré superlattices. *Nature* **597**, 350–354 (2021).
17. Capone, M., Fabrizio, M., Castellani, C. & Tosatti, E. Strongly correlated superconductivity and pseudogap phase near a multiband Mott insulator. *Phys. Rev. Lett.* **93**, 047001 (2004).
18. Nomura, Y., Nakamura, K. & Arita, R. Ab initio derivation of electronic low-energy models for $C_{60}$ and aromatic compounds. *Phys. Rev. B* **85**, 155452 (2012).
19. Nomura, Y., Sakai, S., Capone, M. & Arita, R. Unified understanding of superconductivity and Mott transition in alkali-doped fullerides from first principles. *Sci. Adv.* **1**, e1500568 (2015).
20. Iwahara, N. & Chibotaru, L. F. Dynamic Jahn–Teller instability in metallic fullerides. *Phys. Rev. B* **91**, 035109 (2015).
21. Iwahara, N. & Chibotaru, L. F. Orbital disproportionation of electronic density is a universal feature of alkali-doped fullerides. *Nat. Commun.* **7**, 13093 (2016).
22. Gunnarsson, O. Superconductivity in fullerides. *Rev. Mod. Phys.* **69**, 575 (1997).
23. Suzuki, Y., Shibasaki, S., Kubozono, Y. & Kambe, T. Antiferromagnetic resonance in the Mott insulator fcc-$Cs_3C_{60}$. *J. Phys.: Condens. Matter* 25, 366001 (2013).
24. Kambe, T., Suzuki, Y., Shibasaki, S., Tomita, K. & Kubozono, Y. Superconducting phase diagram in the f.c.c. phase of $Cs_3C_{60}$: Pressure dependence of resistivity. In *Proc. 19th Int. Conf. on Magnetism (ICM2012)* (2012).
25. Terletska, H., Vučičević, J., Tanasković, D. & Dobrosavljević, V. Quantum critical transport near the Mott transition. Phys. Rev. Lett. **107**, 026401 (2011).
26. Hasenbusch, M. Dynamic critical exponent z of the three-dimensional Ising universality class: Monte Carlo simulations of the improved Blume-Capel model. Phys. Rev. E **101**, 022126 (2020).
27. Gibbs, J. W. On the equilibrium of heterogeneous substances. *Trans. Conn. Acad. Arts Sci.* **3**, 108–248 (1876); 343–524 (1878).
28. Potočnik, A. et al. Size and symmetry of the superconducting gap in the f.c.c. $Cs_3C_{60}$ polymorph close to the metal–Mott insulator boundary. *Sci. Rep.* **4**, 4265 (2014).
29. Sun, W., Powell-Palm, M. J. & Chen, J. The geometry of high-dimensional phase diagrams: I. Generalized Gibbs' Phase Rule. arXiv 2105.01337 (2021).
30. Keller, G., Held, K., Eyert, V., Vollhardt, D. & Anisimov, V. I. Electronic structure of paramagnetic $V_2O_3$: Strongly correlated metallic and Mott insulating phase. Phys. Rev. B **70**, 205116 (2004).
31. Ando, Y., Komiya, S., Segawa, K., Ono, S. & Kurita, Y. Electronic phase diagram of high-$T_c$ cuprate superconductors from a mapping of the in-plane resistivity curvature. *Phys. Rev. Lett.* **93**, 267001 (2004).
32. Medvedev, S. et al. Electronic and magnetic phase diagram of β-$Fe_{1.01}Se$ with superconductivity at 36.7 K under pressure. *Nature Mater.* **8**, 630–633 (2009).
33. Li, D. et al. Superconductivity in an infinite-layer nickelate. *Nature* **572**, 624–627 (2019).

34. Timusk, T. & Statt, B. The pseudogap in high-temperature superconductors: An experimental survey. *Rep. Prog. Phys.* **62**, 61 (1999).
35. Mueller, E. J. Review of pseudogaps in strongly interacting Fermi gases. *Rep. Prog. Phys.* **80**, 104401 (2017).
36. Ren, M.-Q. et al. Direct observation of full-gap superconductivity and pseudogap in two-dimensional fullerides. *Phys. Rev. Lett.* **124**, 187001 (2020).
37. Zou, Z. & Anderson, P. W. Neutral fermion, charge-e boson excitations in the resonating-valence-bond state and superconductivity in $La_2CuO_4$-based compounds. *Phys. Rev. B* **37**, 627(1988).
38. Lee, P. A., Nagaosa, N. & Wen, X.-G. Doping a Mott insulator: Physics of high-temperature superconductivity. Rev. Mod. Phys. **78**, 17 (2006).
39. Suhl, H., Matthias, B. T. & Walker, L. R. Bardeen–Cooper–Schrieffer theory of superconductivity in the case of overlapping bands. Phys. Rev. Lett. 3, 552 (1959).
40. Kondo, J. Superconductivity in transition metals. Prog. Theor. Phys. 29, 1–9 (1963).
41. Negele, J. W. & Orland, H. Quantum Many-Particle Systems (CRC Press, Boca Raton, USA, 1998).

## Acknowledgement

This research was supported by the National Natural Science Foundation of China (NSFC Grant Nos. 12174027), Innovation Program for Quantum Science and Technology (2023ZD0300500). Beijing Natural Science Foundation (Grant No. IS25007). This research was also made under the Foreign Researcher Invitation Program of BAQIS. KT thanks Naoya Iwahara, Takahiro Misawa, and Jure Kokalj for their fruitful discussions.

## Author contributions

The research project was organized by KT. The materials were synthesized by YM, SH, XL, and YC. Powder X-ray diffraction and the Rietveld analyses were made by SH. Theoretical consideration on a model Hamiltonian was made by PK. A home-made electrical transport cell was made by YM and SH at AIMR, Tohoku University, Japan. Modifications on an original QD high pressure cell for PPMS, employed for electrical transport measurements at BAQIS, China was made by XL, CY, and HZ. The experimental data collections at early period at Tohoku University, Japan were made by YM and SH. Further additional electrical transport measurements were made by XL, QW, and LZ in BAQIS, China. The interpretations of the experimental data were made by XL, PK, and KT after the final confirmation of all experimental data at AIMR and BAQIS. The manuscript was prepared by XL and confirmed by PK and KT. Final discussions and confirmation were made among all the authors.

## Materials & Correspondence.

Correspondence should be sent to KT, katsumitanigaki@baqis.ac.cn, SH satoshi.heguri@oit.ac.jp, PK kotetes@baqis.ac.cn. Requests for materials and other information can be addressed to KT.

# Supplemental Information

**Supercritical fluid-like metal and insulator coexistent phase and equilibrium Widom line in (Cs, Rb)$_3$C$_{60}$ fulleride superconductors**

Xinying Li[1,2,3], Yuki Matsuda[4], Chongli Yang[1], Huaxue Zhou[1], Takuma Ogasawara[1], Qin Wang[1], Liguo Zhang[1], Hidekazu Shimotani[4], Hailiang Xia[5], Khuong K. Huynh[6], Panagiotis Kotetes[1]*, Satoshi Heguri[7,8]*, Katsumi Tanigaki[1]*

[1]Beijing Academy of Quantum Information Sciences (BAQIS), Bld. #3, No. 10, Xibeiwang East Rd, Haidian District, Beijing, 100193, China

[2] Beijing National Laboratory for Condensed Matter Physics and Institute of Physics, Chinese Academy of Science, 603, Beijing 100190, China

[3]Advanced Institute of Materials Research (AIMR), Tohoku University, 2-1-1 Katahira, Aoba, Sendai, 980-8577, Japan

[4]Material Physics, Institute of Physics, Chinese Academy of Sciences, Beijing 100190, China.

[5]Department of Chemistry, Langelandsgade 140, Building 1512, Room 318, 8000 Aarhus C, Denmark

[6]Osaka Institute of Technology, 16-1-5 Omiya, Asahi, Osaka, Japan

[7]Institute of Physics, Bijenička 46, 10000 Zagreb Croatia

### SI.1 Electrical resistivities (ρ) and magnetic susceptibilities (χ) of Cs$_x$Rb$_{3-x}$C$_{60}$ fullerides.

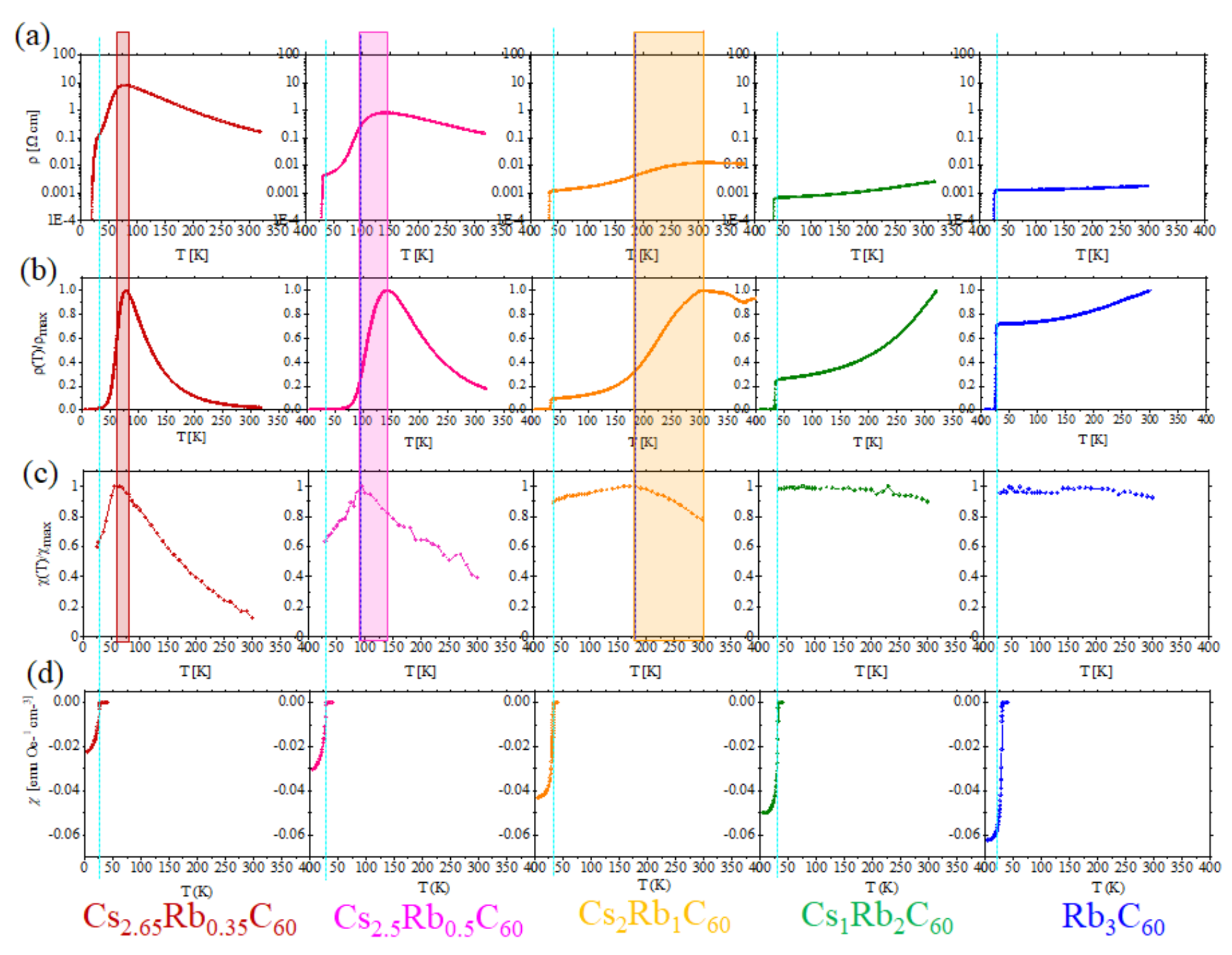

**Fig. S1** Electrical resistivities ((a) logarithmic and (b) linear scale), (c) normal magnetic susceptibilities measured under high magnetic field, and (d) superconducting diamagnetic susceptibilities measured under low magnetic field as a function of temperature of $Cs_xRb_{3-x}C_{60}$ fullerides.

### SI.2 Evolution of $V_{cell}$ parameters as a function of T and P

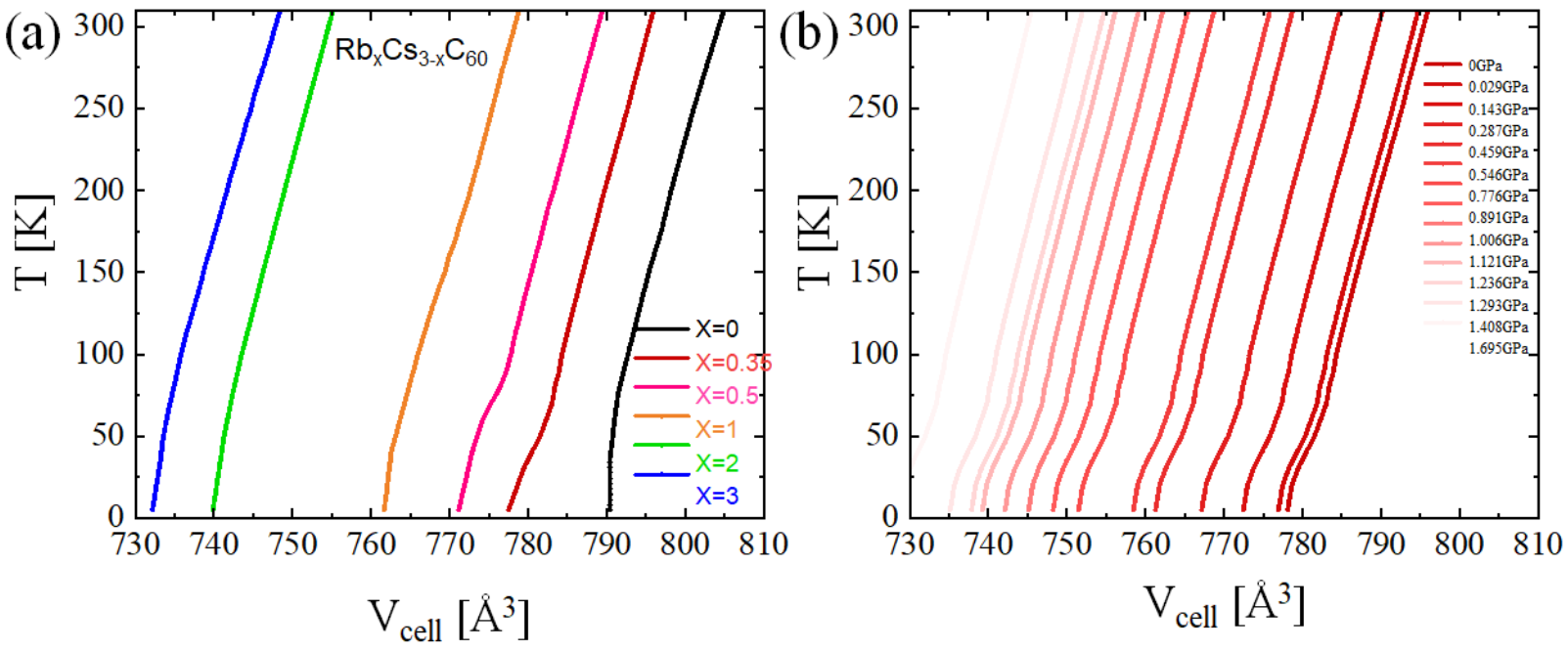

**Fig. S2** The relationships between $V_{cell}$ and temperature (a) $Cs_xRb_{3-x}C_{60}$ (x=0, 0.35, 0.5, 1, 2, 3) with different stoichiometries. (b) $Cs_{2.65}Rb_{0.35}C_{60}$ under various pressures and temperatures. The non-linear pressure dependence of the unit cell volume of $Rb_xCs_{3-x}C_{60}$ fullerides was extracted using the $V_{cell}$ values reported previously[1]. The $V_{cell}$ values were fitted by the semiempirical second-order Murnaghan equation-of-states $P = (K_0/K'_0)\ [(V_0/V)^{K'_0} -1]$, according to the literature using the zero-pressure isothermal bulk moduli ($K_0$) of 19.0, 21.9, 20.0, 20.2, and 20.9 GPa, and pressure derivatives ($K'0$) of 9.0, 8.0, 8.3, 8.0, and 8.8 for x = 0.35, 0.5, 1, and 2, respectively.

### SI.3 Comparison of the magnetic susceptibilities of $Cs_xRb_{3-x}C_{60}$ (x=0, 0.35, 0.5, 1, 2, 3) with previous studies.

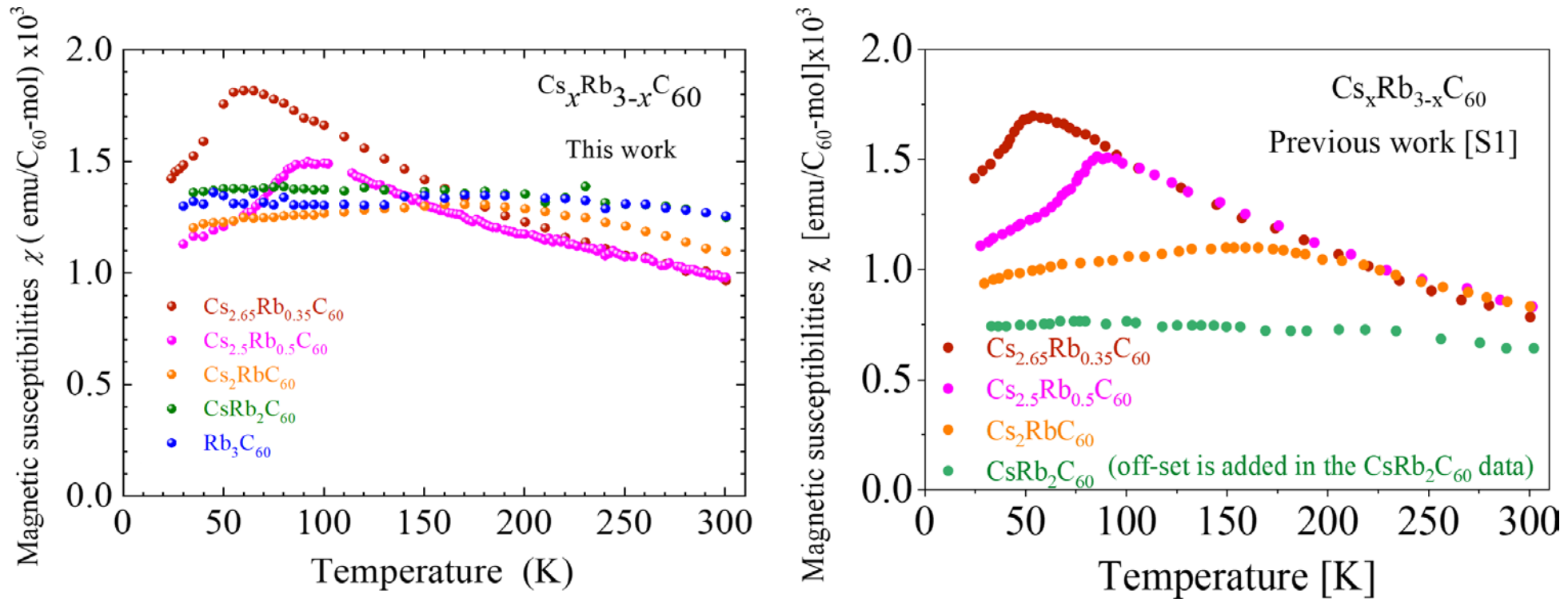

**Fig. S3-1 Magnetic susceptibilities measured under high magnetic fields. The comparison between this work and the previous work is shown.|** Both data sets are qualitatively the same, giving the same MIT line in the V-T phase diagram, but quantitatively show differences from each other. The quantitative difference is essential for making the intrinsic interpretation, as described in the text.

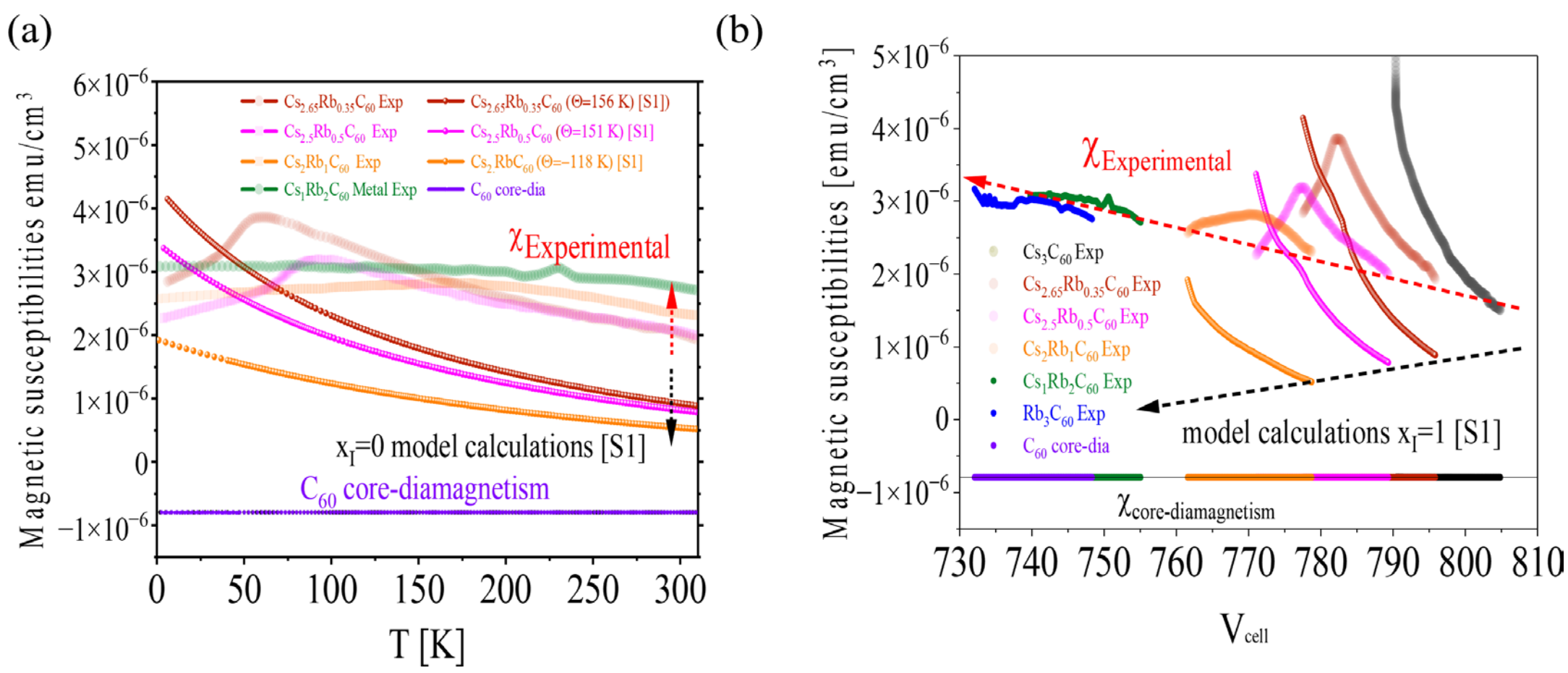

**Fig. S3-2 The magnetic susceptibilities of $Cs_xRb_{3-x}C_{60}$ (x=3, 2.65, 2.5, 2, 1, 0) as a function of $V_{cell}$.|** T evolution of the magnetic susceptibilities of $Cs_xRb_{3-x}C_{60}$ was converted as a function of V(T) by employing the relations between T and V(T) of $Cs_xRb_{3-x}C_{60}$ at ambient pressure in Fig. S2. The magnetic susceptibility of $Cs_3C_{60}$ was taken from the previous literature[1].

**Analysis of magnetic susceptibilities:** the problem of the previous report[1]. In the previous report, an antiferromagnetic Curie-Weiss model was used considering the full insulating state ($x_I = 1$), and the Curie-Weiss temperatures (Θ) were evaluated as (-156 K for $Cs_{2.65}Rb_{0.35}C_{60}$, -151 K for $Cs_{2.5}Rb_{0.5}C_{60}$, and -118 K for $Cs_2RbC_{60}$)[1], by fitting the slopes experimentally observed for $Cs_xRb_{3-x}C_{60}$ fullerides. These values of Θ are large and differ markedly from our theoretical results for $\Theta_{CW}$ calculated using a model Hamiltonian, not only in absolute values but also in a general relationship to $V_{cell}$, as described in Methods. The problem with supposing $x_I = 1$ for analysis is evidently seen in Fig. S3 (a) and (b). Due to this assumption, the Θ values for satisfying the slope of magnetic susceptibilities (χ) were evaluated to be incorrectly too large. Consequently, the resulting χ values become quantitatively too small at finite temperatures, typically at 300 K. As a consequence, a core-diamagnetic susceptibility arising from the hexagonal-ring system of $C_{60}$ π-orbitals (generally a large negative value is experimentally confirmed to be around $-8x10^{-7}$ emu/cm$^3$), has to be inevitably changed to a small value. When this request is beyond the limits of experimental error, a sign of constant value is requested to be positive in sign, as shown in Fig. S4(a). This situation can be observed in $Cs_{2.65}Rb_{0.35}C_{60}$ as reported in the previous literature[1]. From another perspective, the calculated magnetic susceptibilities, based on this method, exhibit an opposite trend to the experimental results. This is significantly different from our experimental data in this work.

As we described in the text, the two MIT lines are drawn based on experiments on σ and χ, and an M-I coexisting pseudo gap (PSG) phase, corresponding to the SCF phase in the G-L phase diagram without symmetry breaking, is emergent in $Cs_xRb_{3-x}C_{60}$ fullerides. In our analysis, we have calculated the Curie-Weiss $\Theta_{CW}(V_{cell})$ values theoretically as a function of $V_{cell}$, based on a model Hamiltonian (see Methods). The computed $\Theta_{CW}(V_{cell})$ values differ greatly from the values evaluated in the previous report[1]. We performed our analyses by employing these theoretical $\Theta_{CW}(V_{cell})$ values without additional hypotheses, while preserving the condition of $x_M + x_I = 1$, and have performed analysis as described in the text.

### SI.4 Electrical conductivities of $Cs_{2.65}Rb_{0.35}C_{60}$ fullerides as functions of T and $V_{cell}$ under pres sure.

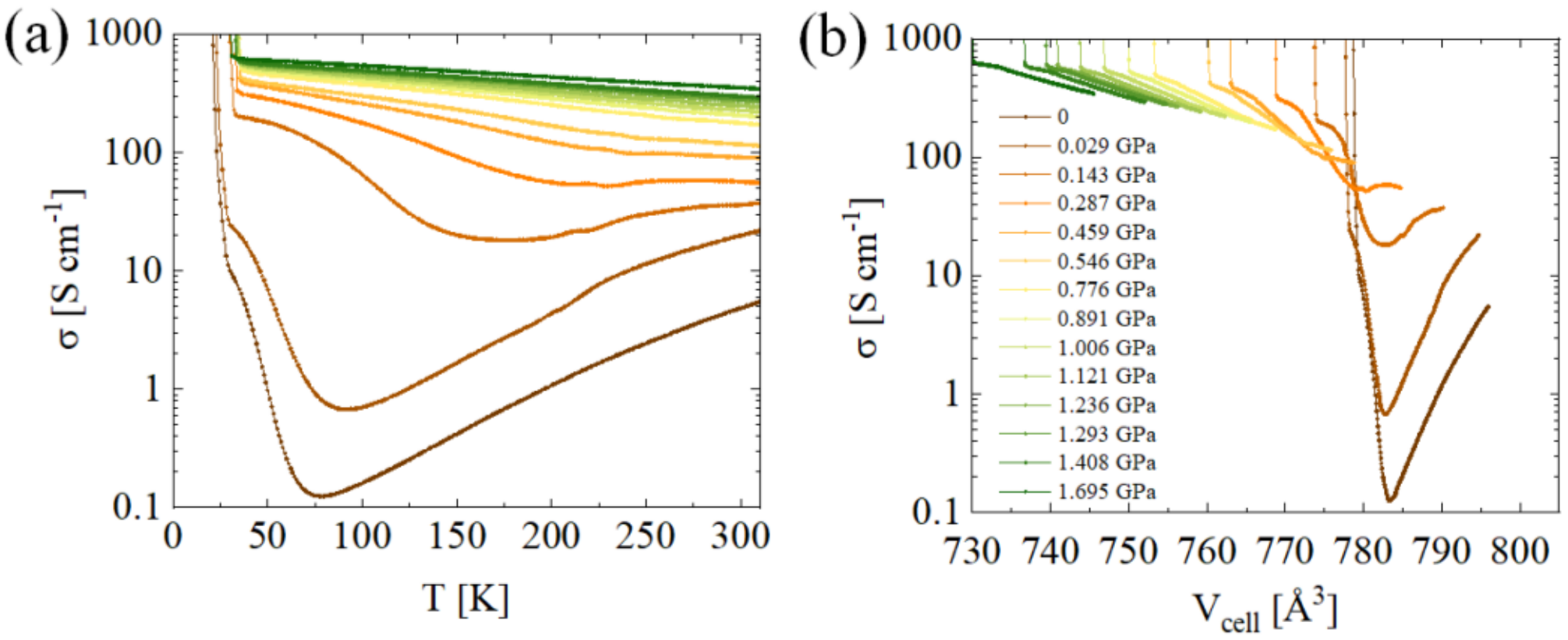

**Fig. S4** Electrical conductivities σ of $Cs_{2.65}Rb_{0.35}C_{60}$ under various P as a function of T (a) and V (b). The $\sigma(V_{cell})$ as a function of T and V. The MIT occurs at around 0.6-0.8 GPa.

### SI.5 Theoretical Analysis: additional information

In the crystal structure of face-centered cubic (fcc) $A_3C_{60}$, there are 12 nearest-neighbor and 6 second- nearest-neighbor $C_{60}$ molecules in a lattice. In a simple two-molecular system, electrons become itinerant via electron transfer from the ground states of $C_{60}^{3-}$ |G:N=3> to $C_{60}^{2-}$ |G; $H_0$ , $H_\pm$: N=2> and $C_{60}^{4-}$ |G; $H_0$, $H_\pm$: N=4> by conserving six electrons in a system. The eigenvectors and eigenvalues, which arise after including the electron transfer Hamiltonian, are found by operating the one-electron transfer Hamiltonian $H_T$ on |G:N=3>, and subsequently diagonalizing the extended Hamiltonian with N = 6. STable.1 provides more detailed information on all the arising $C_{60}^{2-}$ and $C_{60}^{4-}$ excited states.

**Table S1** We enlist all the possible excited states of two decoupled molecules, which are characterized by a total number of N = 6 electrons. These excited states arise when the hopping Hamiltonian $H_T$ (see Methods section) is treated as a perturbation. We provide the absolute energy of each excited state $E_{Excited}$, which is the respective energy relative to the ground state energy $E_G$, along with the corresponding multiplicity (degeneracy) of each excited state. When evaluating the multiplicity, we also account for the $t_{1u}$-orbital degeneracy. In the table, we highlighted the excited state in red, which is energetically closest to the ground state. In the same spirit, we have highlighted in blue the next energetically closest states to the ground state. We observe that all these states constitute spin-singlet states and have a strong contribution to the perturbative result. In the table we introduced the notation: $|1 \text{ pair}; \nu = 0, \pm1\rangle = [\, |\uparrow\downarrow; e; e\rangle + \exp(-i2\pi\nu/3)\, |e; \uparrow\downarrow; e\rangle + \exp(i2\pi\nu/3)|e; e; \uparrow\downarrow\rangle] / 3^{1/2}$ and |2 pairs; ν =

0, ±1> = [ |↑↓; ↑↓; e> + exp(-i2πν/3) |↑↓; e; ↑↓> + exp(i2πν/3)|e; ↑↓; ↑↓>] / $3^{1/2}$.

| State of Molecule $\boldsymbol{R}_i$ (2e) | State of Molecule $\boldsymbol{R}_j$ (4e) | $E_{\text{Excited}}$ | $E_{\text{Excited}} - E_{\text{G}}$ | State Multiplicity |
|---|---|---|---|---|
| $\lvert\uparrow;\uparrow;e\rangle$ | $\lvert\uparrow\downarrow;\uparrow;\uparrow\rangle$ | $7U' - 2J$ | $U' - 6J$ | $3\times3\times3\times3 = 81$ |
| $1/\sqrt{2}(\lvert\uparrow;\downarrow;e\rangle - \lvert\downarrow;\uparrow;e\rangle)$ | $\lvert\uparrow\downarrow;\uparrow;\uparrow\rangle$ | $7U'$ | $U' - 4J$ | $3\times1\times3\times3 = 27$ |
| $\lvert 1\text{ pair}; \nu = 0\rangle$ | $\lvert\uparrow\downarrow;\uparrow;\uparrow\rangle$ | $7U' + 3J$ | $U' - J$ | $1\times1\times3\times3 = 9$ |
| $\lvert 1\text{ pair}; \nu = \pm1\rangle$ | $\lvert\uparrow\downarrow;\uparrow;\uparrow\rangle$ | $7U'$ | $U' - 4J$ | $1\times2\times3\times3 = 18$ |
| $\lvert\uparrow;\uparrow;e\rangle$ | $1/\sqrt{2}(\lvert\uparrow\downarrow;\uparrow;\downarrow\rangle - \lvert\uparrow\downarrow;\downarrow;\uparrow\rangle)$ | $7U'$ | $U' - 4J$ | $3\times3\times3\times1 = 27$ |
| $1/\sqrt{2}(\lvert\uparrow;\downarrow;e\rangle - \lvert\downarrow;\uparrow;e\rangle)$ | $1/\sqrt{2}(\lvert\uparrow\downarrow;\uparrow;\downarrow\rangle - \lvert\uparrow\downarrow;\downarrow;\uparrow\rangle)$ | $7U' + 2J$ | $U' - 2J$ | $3\times1\times3\times1 = 9$ |
| $\lvert 1\text{ pair}; \nu = 0\rangle$ | $1/\sqrt{2}(\lvert\uparrow\downarrow;\uparrow;\downarrow\rangle - \lvert\uparrow\downarrow;\downarrow;\uparrow\rangle)$ | $7U' + 5J$ | $U' + J$ | $1\times1\times3\times1 = 3$ |
| $\lvert 1\text{ pair}; \nu = \pm1\rangle$ | $1/\sqrt{2}(\lvert\uparrow\downarrow;\uparrow;\downarrow\rangle - \lvert\uparrow\downarrow;\downarrow;\uparrow\rangle)$ | $7U' + 2J$ | $U' - 2J$ | $1\times2\times3\times1 = 6$ |
| $\lvert\uparrow;\uparrow;e\rangle$ | $\lvert 2\text{ pairs}; \nu = 0\rangle$ | $7U' + 3J$ | $U' - J$ | $3\times3\times1\times1 = 9$ |
| $1/\sqrt{2}(\lvert\uparrow;\downarrow;e\rangle - \lvert\downarrow;\uparrow;e\rangle)$ | $\lvert 2\text{ pairs}; \nu = 0\rangle$ | $7U' + 5J$ | $U' + J$ | $3\times1\times1\times1 = 3$ |
| $\lvert 1\text{ pair}; \nu = 0\rangle$ | $\lvert 2\text{ pairs}; \nu = 0\rangle$ | $7U' + 8J$ | $U' + 4J$ | $1\times1\times1\times1 = 1$ |
| $\lvert 1\text{ pair}; \nu = \pm1\rangle$ | $\lvert 2\text{ pairs}; \nu = 0\rangle$ | $7U' + 5J$ | $U' + J$ | $1\times2\times1\times1 = 2$ |
| $\lvert\uparrow;\uparrow;e\rangle$ | $\lvert 2\text{ pairs}; \nu = \pm1\rangle$ | $7U'$ | $U' - 4J$ | $3\times3\times1\times2 = 18$ |
| $1/\sqrt{2}(\lvert\uparrow;\downarrow;e\rangle - \lvert\downarrow;\uparrow;e\rangle)$ | $\lvert 2\text{ pairs}; \nu = \pm1\rangle$ | $7U' + 2J$ | $U' - 2J$ | $3\times1\times1\times2 = 6$ |
| $\lvert 1\text{ pair}; \nu = 0\rangle$ | $\lvert 2\text{ pairs}; \nu = \pm1\rangle$ | $7U' + 5J$ | $U' + J$ | $1\times1\times1\times2 = 2$ |
| $\lvert 1\text{ pair}; \nu = \pm1\rangle$ | $\lvert 2\text{ pairs}; \nu = \pm1\rangle$ | $7U' + 2J$ | $U' - 2J$ | $1\times2\times1\times2 = 4$ |

**SI.6 Slope of the 1st-order MIT line based on Clapeyron-Clausius analyses**

The Gibbs free energy (G) is described in terms of two intensive parameters of T and P. When the G of a material decreases, a spontaneous change proceeds according to the Gibbs energy. In the present paper, we describe our experimental results of electrical transport and magnetic susceptibility as a T-$V_{cell}$(T, P) phase diagram. Since $V_{cell}$ denotes a cell volume per one $C_{60}$ molecule, $V_{cell}$(T, P) is a quantity of an intensive parameter in place of P(T, V). The Clausius-Clapeyron law $(\partial P/\partial T)_{MIT} = \Delta S/\Delta V$ satisfies the equilibrium between the insulator (I) and the metal (M) phase at the same chemical potential, μ(I) = μ(M). The equation connects the macroscopic quantity of a volume change (ΔV) with a microscopic entropy change (ΔS) derived from Boltzmann's formula $S = k_B \ln W_{BZ}$ in the phase equilibrium, where $W_{BZ}$ represents the number of accessible quantum states and $k_B$ is the Boltzmann constant. The evolution of $(\partial T/\partial P)_{MIT}$ as a function of P was evaluated to be 65 ～ 15 K/kbar as shown in Fig. S6(a), based on the T-P phase diagram of Fig.4(b), as explained in the text. ΔS can be evaluated by $\Delta S = [\Delta V/(\partial T/\partial P)_{MIT}]\text{x}(10^2/4.184)$, where $10^2/4.184$ is the unit conversion factor for expressing $(\partial P/\partial T)_{MIT}$ with the unit of K/kbar. Based on the electrical conductivity (σ) of $Cs_{2.65}Rb_{0.35}C_{60}$ as a function of P shown in Fig. S5(b), the MIT occurs at around P = 0.6 - 0.8 GPa. The ΔV value of $Cs_{2.65}Rb_{0,35}C_{60}$ during MIT is $5Å^3/V_{cell}(A_3C_{60})$ = 5 x$10^{24}$ $cm^3/C_{60}$-mole = 3.01 $cm^3/C_{60}$-mol[1]. Given the linear response of V(P) as shown in Fig. S6(b), we evaluated the evolution of ΔS as a function of P, and its value increases from 1.1 to 4.8 with a plateau at ΔS = 2.5 cal/K at P = 6.5 kbar, as shown in Fig. S6(c).

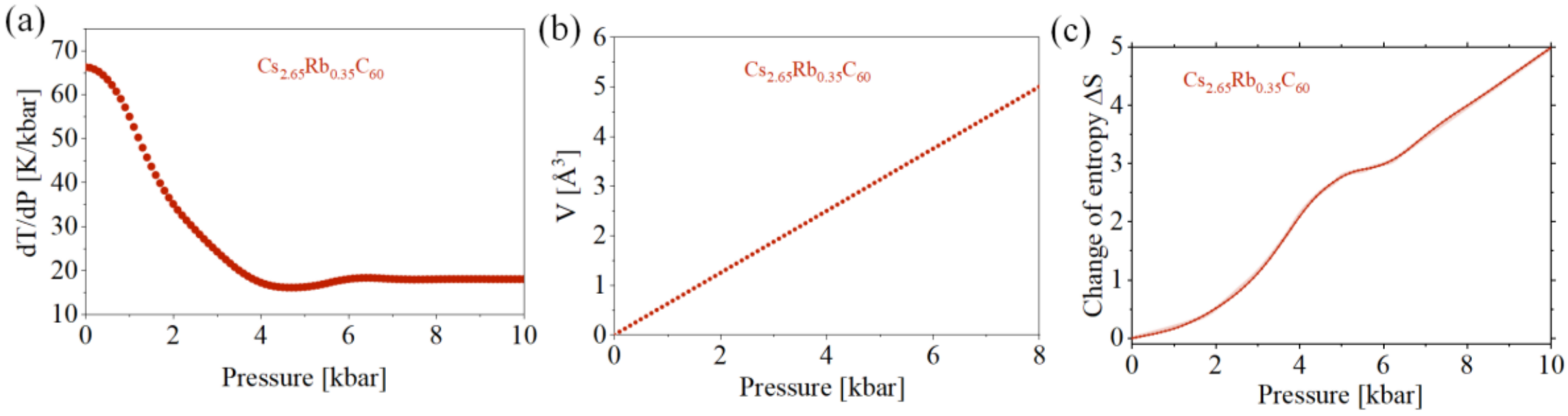

**Fig. S6** (a) The $(dT/dP)_{MIT}$ calculated from the T-P phase diagram for $Cs_{2.65}Rb_{0.35}C_{60}$ shown in Fig.4(b). (b) Linear response of V(P) as a function of P during MIT. MIT occurs P = 6 - 8 kbar according to the conductivity in Fig. S5(a, b). (c) Pressure dependence of the calculated entropy change, following the Clausius-Clapeyron law $(\partial P/\partial T)_{MIT} = \Delta S/\Delta V$.

According to the model Hamiltonian of $A_3C_{60}$ (see the Method section) and SI.6, the microstate degeneracy W associated with each electronic configuration is determined by the orbital degeneracy of the $t_{1u}$ manifold under the dynamic Jahn–Teller effect. In the Mott insulating regime, electron transfer induced by the transfer integral t connects the decoupled ground state |G, N=3>⊗|G, N=3> with the charge-transferred state |ν=0:N=2>⊗|ν=0:N=4>. The corresponding entropy is evaluated as $\Delta S = N_A k_B \ln W \simeq k_B N_A \ln(3)$ = 2.2 cal/(K $C_{60}$-mol), based on the Boltzmann law where $k_B$ is the Boltzmann constant and $N_A$ is Avogadro's number. The theoretically evaluated value based on the model Hamiltonian is in fairly good agreement with the experimental observation.

### SI.7 Electrical conductivities and magnetic susceptibilities for $A_3C_{60}$ fullerides, κ-Cu[(BEDT-TTF)$(CN)_2$]Cl, cuprate (LSCO), and 2-dimensional stacked layer material of $MoTe_2/WSe_2$.

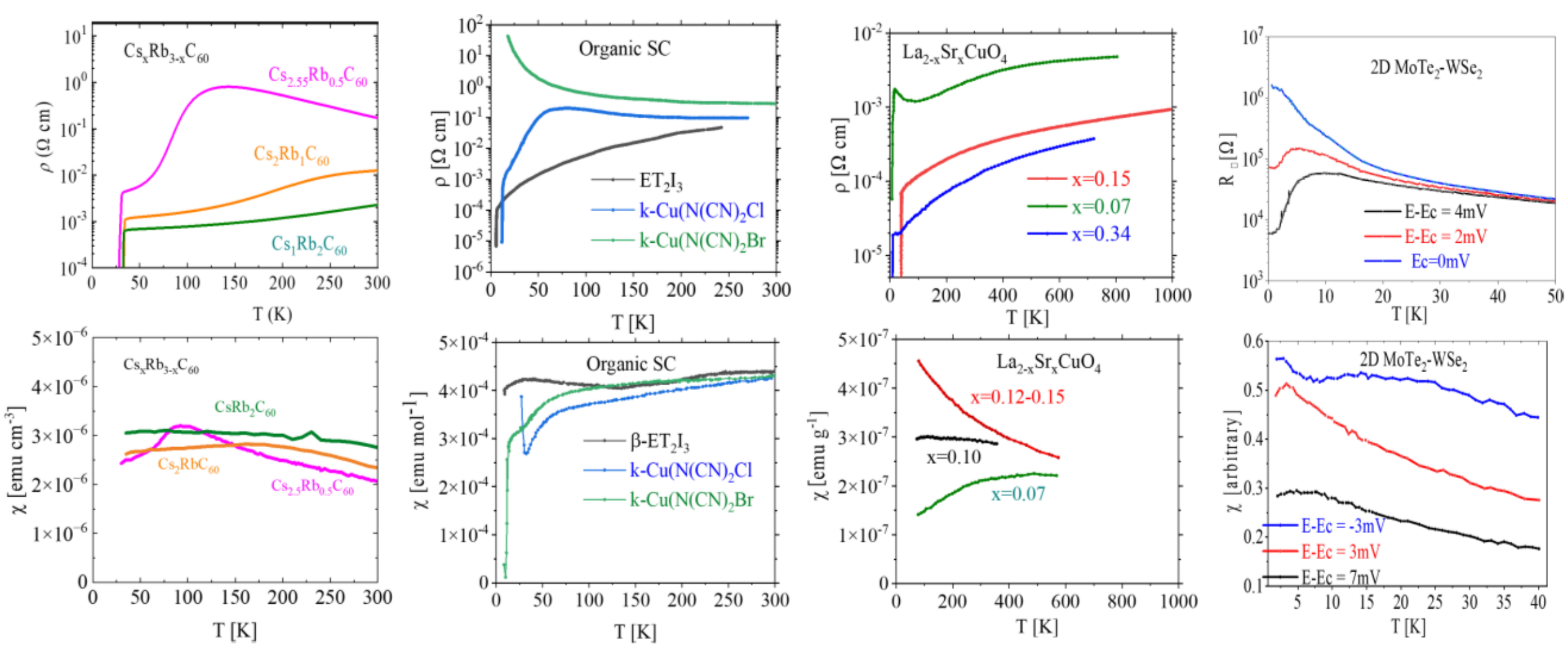

**Fig. S7** Resistivities ρ and magnetic susceptibilities χ as a function of temperature. The comparisons are made among $Cs_xRb_{3-x}C_{60}$ fullerides, organic superconductors, LSCO cuprate superconductors, and two-dimensional moirè $MoTe_2/WSe_2$ materials. The data are collected from the literatures[2-6].

### SI.8 Electrical resistivities at pressures around the formation in SC $T_c$ dome.

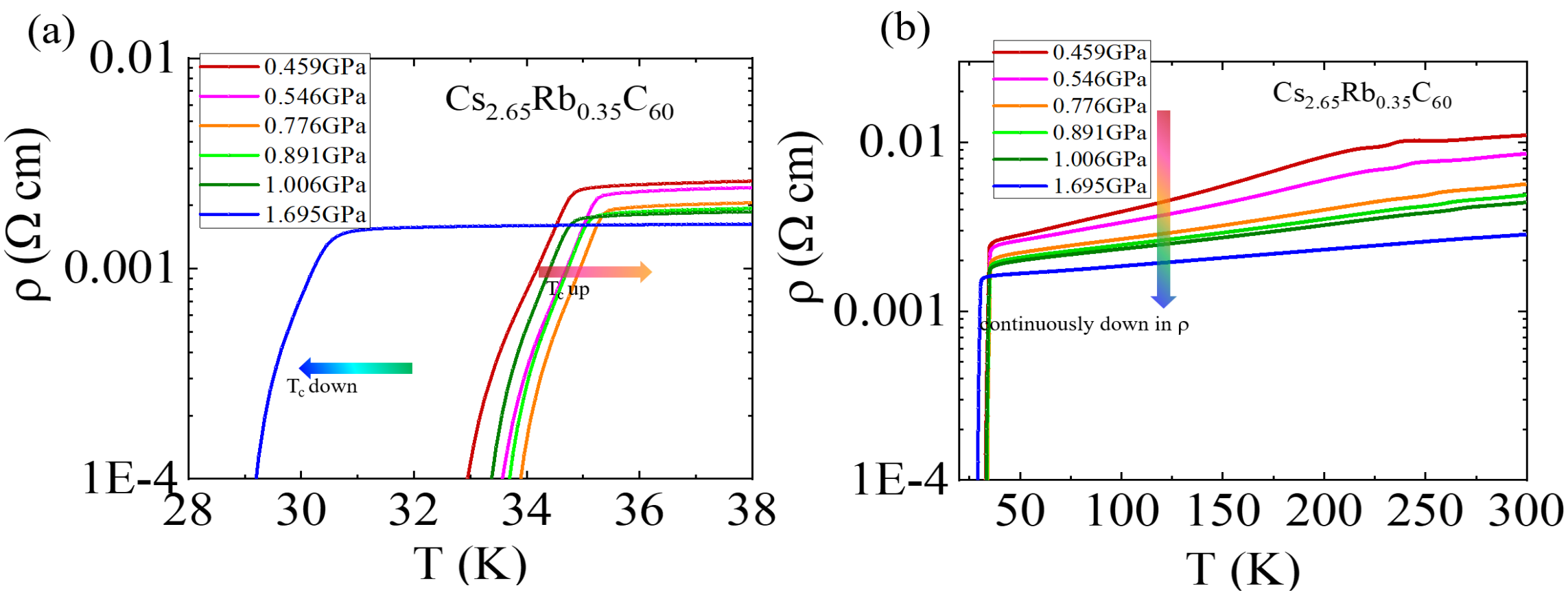

**Fig. S8** (a) Resistivity (ρ) observed for $Cs_{2.65}Rb_{0.35}C_{60}$ fulleride under various pressures. The $T_c$ rises up first and then decreases, displaying a SC $T_c$ dome. The value of ρ decreases continuously with $V_{cell}$ without anomalies, despite the formation of the SC $T_c$-dome.

### SI.9 Scaling based on electrical resistivities of $Cs_{2.65}Rb_{0.35}C_{60}$ as a function of $V_{cell}$.

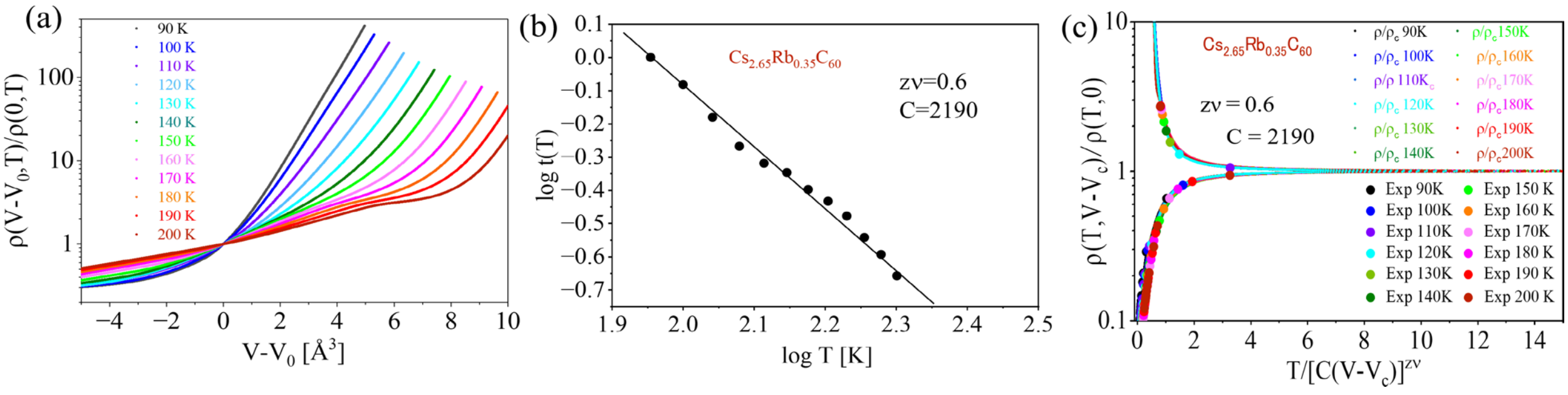

**Fig. S9** (a) The relationship between $\rho(V\text{-}V_0,T)/\rho_0(0,T)$ and $V\text{-}V_0$ obtained from Fig. S2(b) and Fig. S5(a,b). (b) The relationship between log(t(T)) and logT for evaluating the critical component values $z\nu$ and C in scaling parameters. (c) Scaling between $\rho(V\text{-}V_0,T)/\rho_0(0,T)$ and $F[T/(C(V\text{-}V_0))^{z\nu}]$, where F is the scaling function. The scaling analysis was performed using the same analytical method as described in the previous literature[7,8], in which the scaling function F is $\exp[\pm(T/T_0)^{-1/z\nu}]$. The final scaling picture (Fig. 4(a)) was produced according to the procedure. The critical exponent $z\nu = 0.6$ with C=2190, being in similar values to those of organic conductors[8]. The critical exponents belong to the exponents in the quantum regime predicted by theory[8], and apart from the classical Ising-type exponent values[9].

### References of SI

1. Zadik, R. H. et al. Optimized unconventional superconductivity in a molecular Jahn-Teller metal. Sci. Adv. 1, e1500059 (2015).

2. Kanoda, K. Metal‑insulator transition in κ-(ET)2X and (DCNQI)2M: Two contrasting manifestations of electron correlation. J. Phys. Soc. Jpn. 75, 051007 (2006).
3. Li, T. et al. Continuous Mott transition in semiconductor moiré superlattices. Nature 597, 350‑354 (2021).
4. Oda, M., Ohguro, T., Yanada, N. & Ido, M. Magnetic susceptibility and superconductivity in $La_{1-x}Sr_xCuO_4$. *J. Phys. Soc. Jpn.* **58**, 1137 (1989).
5. Kastner, M. A., Birgeneau, R. J., Shirane, G. & Endoh, Y. Magnetic, transport, and optical properties of monolayer copper oxides. *Rev. Mod. Phys.* **70**, 897 (1998).
6. Ando, Y., Lavrov, A. N., Komiya, S., Segawa, K. & Sun, X. F. Mobility of the doped holes and the antiferromagnetic correlations in underdoped high-$T_c$ cuprates. *Phys. Rev. Lett.* **87**, 017001 (2001).
7. Kagawa, F., Miyagawa, K. & Kanoda, K. Magnetic Mott criticality in a κ-type organic salt probed by NMR. Nat. Phys. 5, 880‑884 (2009).
8. Terletska, H., Vučičević, J., Tanasković, D. & Dobrosavljević, V. Quantum critical transport near the Mott transition. Phys. Rev. Lett. 107, 026401 (2011).
9. Hasenbusch, M. Dynamic critical exponent z of the three-dimensional Ising universality class: Monte Carlo simulations of the improved Blume-Capel model. Phys. Rev. E 101, 022126 (2020).